\documentclass[varenna]{cimento}
\usepackage{amsfonts}
\usepackage{amsmath}
\usepackage[british]{babel}
\usepackage{graphicx}
\usepackage{amssymb}

\newcommand{\itT}{\textit{T}}

\newcommand{\itE}{\textit{E}}

\newcommand{\itTpE}{\textit{T+E}}
\newcommand{\Planck} {\textit{Planck }}

\def\Commander{\texttt{Commander}}

\def\NILC{\texttt{NILC}}
\def\SMICA{\texttt{SMICA}}
\def\SEVEM{\texttt{SEVEM}}
\newbox\tablebox    \newdimen\tablewidth
\def\leaderfil{\leaders\hbox to 5pt{\hss.\hss}\hfil}
\def\endPlancktable{\tablewidth=\columnwidth 
    $$\hss\copy\tablebox\hss$$
    \vskip-\lastskip\vskip -2pt}

\def\tablenote#1 #2\par{\begingroup \parindent=0.8em
    \abovedisplayshortskip=0pt\belowdisplayshortskip=0pt
    \noindent
    $$\hss\vbox{\hsize\tablewidth \hangindent=\parindent \hangafter=1 \noindent
    \hbox to \parindent{$^#1$\hss}\strut#2\strut\par}\hss$$
    \endgroup}
\def\doubleline{\vskip 3pt\hrule \vskip 1.5pt \hrule \vskip 5pt}

\title{Primordial Non-Gaussianity}

\author{M. Celoria}

\institute{
 Gran Sasso Science Institute, via F. Crispi 7, I-67100, L'Aquila, Italy
 \\
 ICTP, International Centre for Theoretical Physics
Strada Costiera 11, 34151, Trieste, Italy
}

\author{S. Matarrese}
\institute{
Dipartimento di Fisica e Astronomia ``Galileo Galilei'',  Universit\`a degli Studi di Padova,
\\
 via Marzolo 8, I-35131, Padova, Italy
\\
INFN, Sezione di Padova, via Marzolo 8, I-35131, Padova, Italy 
\\
INAF-Osservatorio Astronomico di Padova, vicolo dell Osservatorio 5, I-35122 Padova, Italy 
\\
Gran Sasso Science Institute, via F. Crispi 7, I-67100, L'Aquila, Italy }

\shortauthor{M. Celoria \atque S. Matarrese}

\begin{document}

\maketitle

\begin{abstract}
Here we review the present status of modelling of and searching for primordial non-Gaussianity of cosmological perturbations. After introducing the models for non-Gaussianity generation during inflation, we discuss the search for non-Gaussian signatures in the Cosmic Microwave Background and in the Large-Scale Structure of the Universe. 
\end{abstract}

\section{Introduction}
According to {\it Planck} 2018, ``the 6-parameter $\Lambda$CDM model provides an astonishingly accurate description of the Universe form times prior to ~380,000  years after the Big Bang, defining the last-scattering surface observed via the Cosmic Microwave Background (CMB) radiation, to the present day at an age of 13.8 billion years" \cite{Aghanim:2018eyx}.
\\
Actually, the concordance model describes the evolution  of tiny fluctuations on top of a homogeneous and isotropic background from the early Universe to the time of observation.
Specifically, we can study how these modes looked like  at the last-scattering surface, by analyzing the inhomogeneities in the CMB, and, in recent times, by observing  the Large-Scale Structure (LSS) of the Universe.

In the standard model of cosmology, the primordial perturbations, corresponding to the  seeds for the LSS,  are chosen  from a Gaussian distribution with random phases. This assumption is justified from experimental evidences as deviations from this Gaussian hypothesis, i.e. Primordial Non-Gaussianity (PNG), have not been observed yet.  
\\
Note that from the theoretical point of view, it is not surprising that ``a Gaussian random field may provide a good description of the properties of density fluctuations'' \cite{Bardeen:1985tr}.
Actually, ``the central limit theorem implies that a Gaussian distribution arises whenever one has a variable [...] which is a linear superposition of a large number of independent random variables  [...] which are all drawn from the same distribution''  \cite{Bardeen:1985tr}. 
\\
For this reason, deviations from perfect Gaussianity can provide relevant information on the early Universe and research on PNG is particularly important, especially if these initial conditions were generated by some dynamical process, such as, for example, inflation in the Early Universe.
\\
Actually,  while ``small-amplitude curvature perturbations generated by quantum fluctuations in an inflationary phase [...] would yield a nearly Gaussian random density field'' \cite{Bardeen:1985tr}, direct measurements of non-Gaussianity would allow us to go beyond the free-field limit, providing information concerning the degrees of freedom, the possible symmetries and the interactions characterizing   the inflationary action.

\subsection{Historical outline}

To be, or not to be Gaussian? The quest for Non-Gaussianity  (NG) has a long story, already during the late seventies observations indicated that the patterns in the LSS  could not be related to a Gaussian distribution. 
More precisely, NG in the LSS was measured in 1977 by Groth and Peebles \cite{Groth:1977gj} who computed the 3-point function of galaxies, raising the question whether this  feature was only associated to non-linear gravitational clustering or it also included some signature of primordial NG.

In the subsequent years, and especially during the late eighties,  the consequences of strongly non-Gaussian initial conditions were investigated in order to explore alternative structure formation models.
However, these extreme possibilities were later excluded by   CMB and LSS observations with increased accuracy.

\begin{figure}
  \centering
    \includegraphics[width=0.9\textwidth]{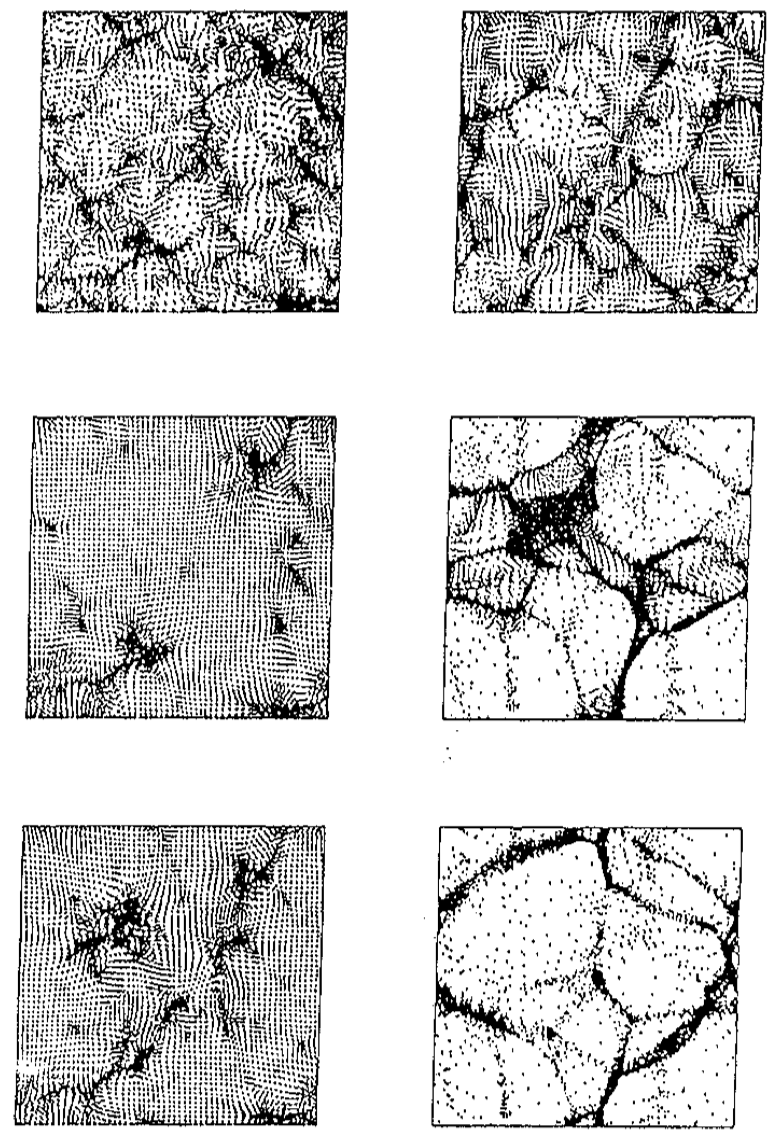}
    \caption{Projected particle positions in slices of depth one sixteenth of the computational box-size at the present time $t_0$. The slices refer to different models. From \cite{Moscardini:1991hf}.}
\label{viewNG1990}
\end{figure}

Remarkably, the early nineties featured the beginning of a new era of non-Gaussian models from inflation, characterized by a small  $f_{\rm NL}$, compatible with observations \cite{Salopek:1990jq, Salopek:1990re, Gangui:1993tt, Verde:1999ij, Komatsu:2001rj}. 
At the same time, N-body simulations started to play a crucial role determining the  LSS of the Universe arising from the non-linear gravitational clustering of non-Gaussian Cold Dark Matter perturbations. The view on NG using N-body simulations around 1990 can be depicted in Figure \ref{viewNG1990}.
From the theoretical point of view, this line of research continued until  the new millennium, when PNG finally emerged as a new ``smoking gun'' of (non-standard)  inflation models \cite{Acquaviva:2002ud, Maldacena:2002vr},  probing  interactions among fields at the highest energy scales, which complements the search for primordial gravitational-waves (PGW). 

As for the experimental and observational side,    the  bispectra for PSCz\cite{Feldman:2000vk}  and the IRAS \cite{Scoccimarro:2000sp} redshift catalogues were determined in 2001, and   for 2dF galaxies   in 2002 \cite{Verde:2001sf}. More recently,   the three-point correlation functions for the WiggleZ spectroscopic galaxy survey and  the Baryon Oscillation Spectroscopic Survey were determined in   \cite{Marin:2013bbb}  and \cite{Gil-Marin:2014sta, Gil-Marin:2016wya},  respectively. 
\footnote{
Note that even if the sensitivity is not competitive with CMB data \cite{Verde:1999ij}, interesting bounds on local $f_{\rm NL}$ from current power spectrum constraints can be found in \cite{Padmanabhan:2006cia, Slosar:2008hx, Xia:2010yu, Xia:2011hj, Nikoloudakis:2012fa, Agarwal:2013qta, Karagiannis:2013xea, Leistedt:2014zqa}.
}

Finally, the very stringent {\it Planck} constraints on PNG \cite{Ade:2013ydc, Ade:2015ava} rose the question whether this route is still viable or not, and the present-day challenge is  to detect (or constrain) mild or weak deviations from primordial Gaussian initial conditions. 

\section{Non-Gaussianity in the Initial Conditions}

The inflationary paradigm, a phase of accelerated expansion in the early Universe, was originally proposed at the end of the seventies in order to overcome some inconsistencies of the hot Big Bang model, which was plagued by the so-called flatness and   horizon problems.
At the same time, inflation suggested a quantum origin for the density fluctuations in the Universe, thereby providing a convincing dynamical mechanism for structure formation.
Generally, the testable predictions of  inflationary models are
\begin{itemize}
\item a critical value for the total energy density;
\item almost, but not exact, scale-invariant  and nearly Gaussian adiabatic density fluctuations;
\item almost, but not exact, scale-invariant stochastic background of relic gravitational waves.
\end{itemize}
Note that {\it Planck} data have confirmed these predictions, for example the measured spectral index of the scalar power spectrum is  $n_s=0.9649\pm 0.0042$
 at 68\% CL and  no evidence for a scale dependence of $n_s$ has been found \cite{Akrami:2018odb}.
 \\
Also spatial flatness is confirmed at a precision of 0.4\% at 95\% CL with the combination of BAO data \cite{Akrami:2018odb}.  Prospects for further improving measurements of spatial curvature are discussed in \cite{Jimenez:2017tkk}.
\\
While primordial gravitational waves have not been yet detected, 
 the upper limit on the tensor-to-scalar ratio
from the BICEP2/Keck CMB polarization experiments  is $r_{0.05} < 0.07$ at 95\% confidence, which tightens to $r_{0.05} < 0.06$ in conjunction with {\it Planck} temperature measurements and other data  \cite{Akrami:2018odb, Ade:2018gkx}. 

In order to reconstruct the inflationary action, we need two ingredients:
\begin{itemize}
\item  the stochastic GW background, providing information on the inflationary energy scales;
\item  deviations from Gaussian initial conditions, providing information on the possible interactions. Moreover, PNG features can help us to distinguish inflation models which would yield the same predictions for $n_s$ and $r$.
\end{itemize}
Many primordial (inflationary) models of non-Gaussianity can be represented in configuration space by the simple formula \cite{Salopek:1990jq, Salopek:1990re, Gangui:1993tt, Verde:1999ij, Komatsu:2001rj}
\begin{equation}
\Phi=\varphi_L+f_{\rm NL} \left( \varphi_L^2-\left<\varphi_L^2\right>\right) +g_{\rm NL} \left(\varphi_L^3-\left<\varphi_L^2\right>\varphi_L\right)+\dots
\end{equation}
where $\Phi$ is the large-scale gravitational potential (or equivalently in terms of the  gauge-invariant comoving curvature perturbation $\zeta$ which on super-horizon scales satisfies the relation $\Phi = 3\, \zeta/5 $), $\varphi_L$ its linear Gaussian contribution and $f_{\rm NL}$ the dimensionless non-linearity parameter (or more generally non-linearity function). 

\subsection{Non-Gaussianity and higher-order statistics}

The simplest statistics measuring NG is the 3-point function or its Fourier transform, the ``bispectrum'':
\begin{equation}
 \left< \Phi(\mathbf{k}_1) \Phi(\mathbf{k}_2)\Phi(\mathbf{k}_3)\right> = (2\pi)^3\delta^{(3)}(\mathbf{k}_1+\mathbf{k}_2+\mathbf{k}_3) B_\Phi(k_1,k_2,k_3)
\end{equation}
which carries shape information.
In the simple linear and quadratic model, parametrized by $\varphi_L$ and $f_{\rm NL}$, the bispectrum of the gravitational potential reads
\begin{equation}
 B_\Phi(k_1,k_2,k_3)= 2f_{\rm NL} [P_\Phi(k_1)P_\Phi(k_1) + \text{cyclic terms}]
\end{equation}
where we applied the Wick's theorem and
\begin{equation}
\left<\Phi(\mathbf{k}_1)\Phi(\mathbf{k}_2)\right> = (2\pi)^3\delta^{(3)}(\mathbf{k}_1+\mathbf{k}_2) P_\Phi(k_1) \, .
\end{equation}   
In order to evaluate NG from the early Universe to the present time, we need, first of all,  to calculate non-Gaussianity during inflation using a self-consistent method.
\\
Then, we need to evolve scalar (vector) and tensor perturbations to second order  outside the horizon, matching conserved second-order gauge-invariant variables, such as the comoving curvature perturbation $\zeta^{(2)}$ (or non-linear generalizations of it), to its value at the end of inflation (accurately accounting for reheating).
\\
Finally, we  consistently study the evolution of  the perturbations  after they re-entered the Hubble radius, by computing the second-order radiation transfer function for the CMB and the second-order matter transfer function for the LSS.

Although this procedure is  involved,  PNG represents a fundamental tool to  probe fundamental physics (e.g. UV completion of the standard model of particle physics or general relativity such as string theory) during inflation at energies from the  Grand Unified Theories (GUT) scale $\sim 10^{15}$ GeV to the Planck scale $\sim 10^{19}$ GeV, as different inflationary models predict different amplitudes and shapes of the bispectrum.
For example, even tough standard models of slow-roll inflation predict  tiny deviations from Gaussianity \cite{Salopek:1990jq, Salopek:1990re, Gangui:1993tt, Verde:1999ij, Komatsu:2001rj,  Acquaviva:2002ud, Maldacena:2002vr}, consistent with the 2013 and 2015 {\it Planck} results, specific oscillatory PNG features can point to  particular  string-theory models as shown in \cite{Arkani-Hamed:2015bza, Silverstein:2017zfk}.

In conclusion,  searching for PNG is interesting per-se for theoretically well-motivated models of inflation and, as shown in {\it Planck} 2015 results \cite{Ade:2015ava} (see also \cite{Akrami:2018odb}), can severely limit various classes of inflationary models beyond the simplest paradigm.

\subsection{Bispectrum of a self-interacting scalar field in de Sitter space}

Consider a scalar field $\chi$ with cubic self-interactions, i.e. with an interaction term of the form $\lambda \, \chi^3/6$ in
the Lagrangian. 
\\
Writing the field as $\chi=\chi_0+\delta\chi$, where $\delta\chi$ represent the fluctuations around its vacuum expectation value $\chi_0=\left<0|\chi|0\right>$, the two and  three-point functions in Fourier space \cite{Falk:1992sf, Gangui:1993tt, Bernardeau:2004zz}, after the rescaling  $\delta\chi=\delta\hat\chi/a$  are given by
\begin{equation}
\begin{aligned}
\left<0| \delta\hat\chi(\tau, \mathbf{k})\delta\hat\chi(\tau', \mathbf{k}')|0\right>
&=
\delta^{(3)}(\mathbf{k}+\mathbf{k}') G(k, \tau, \tau')
\\
\left< \delta\hat\chi_{\mathbf{k}_1}\delta\hat\chi_{\mathbf{k}_2}\delta\hat\chi_{\mathbf{k}_3}
\right>
&=i\, \lambda\, \delta^{(3)}(\mathbf{k}_1+\mathbf{k}_2+\mathbf{k}_3)\times
\\
&\times
\int_{-\infty}^\tau\frac{d\tau'}{H\tau'}
\left[\prod_{i=1}^3G(k_i, \tau,\tau')
-\prod_{i=1}^3G^\ast(k_i, \tau,\tau')\right]
\end{aligned}
\end{equation}
where the Green's function reads
\begin{equation}
G(k_i, \tau,\tau')=\frac{1}{2k_i}\left( 1-\frac{i}{k_i\tau}\right)\left( 1+\frac{i}{k_i\tau'}\right)e^{ik_i(\tau'-\tau)} \, .
\end{equation}
The bispectrum is ($\zeta$ being a function of order $1$) 
\begin{equation}
\left< \delta\chi_{\mathbf{k}_1}\delta\chi_{\mathbf{k}_2}\delta\chi_{\mathbf{k}_3}
\right>
=\sum_i\nu_3(k_i)\prod_{j\neq i}\frac{H^2}{2k_j^3}
\end{equation}
where
\begin{equation}
\nu_3(k_i)=\frac{\lambda}{3H^2}\left[  \gamma +\zeta(k_i)+\log(-k_T\tau)  \right] \, .
\end{equation}
Historically, in \cite{Falk:1992sf} it was  found $f_{\rm NL}\sim \epsilon^2$  for the standard single-field slow-roll scenario (from non-linearity in the inflaton potential in a fixed de Sitter space-time).
\\
Later,  calculations from second-order gravitational corrections during stochastic inflation   indicated  $f_{\rm NL} \sim  \epsilon, \eta$ \cite{Gangui:1993tt}. 
This result  has been confirmed in \cite{Acquaviva:2002ud, Maldacena:2002vr}, up to numerical factors and momentum-dependent terms, with a full second-order approach. 
\\
Finally, Weinberg extended the calculation of the bispectrum to 1-loop \cite{Weinberg:2005vy}. 
Remarkably, one of  the terms gives rise to the so-called ``consistency relation'', according to which  $f_{\rm NL} = - 5\, (n_s-1)/12$.
\\
However,  it has been shown that the ``consistency relation'' term can be gauged away by a non-linear rescaling of coordinates, up to sub-leading terms. Hence the only residual term is proportional to $\epsilon$ i.e. to the amplitude of tensor modes; see  comments on this point, later on.

\subsection{Shapes of non-Gaussianity from inflation}

In order to extract the relevant  information regarding the amplitude and shape of PNG, it is convenient to write the bispectrum of primordial curvature perturbations as
\begin{equation}
\left<\zeta(\mathbf{k}_1)\zeta(\mathbf{k}_2)\zeta(\mathbf{k}_3)
\right>
=
(2\pi)^3\delta^{(3)}(\mathbf{k}_1+\mathbf{k}_2+\mathbf{k}_3)f_{\rm NL}F(k_1,k_2,k_3)
\end{equation}
where $f_{\rm NL}$ represents the amplitude, while $F(k_1,k_2,k_3)$ encodes the shape of PNG.
Note that, usually, we study the function $F(1,x_2,x_3)x_2^2x^2_3$ in terms of the rescaled coordinates $x_2=k_2/k_1$ and $x_3=k_3/k_1$, where momenta  satisfy the triangle inequality $x_2 + x_3 > 1$.

Remarkably, there are several possible shapes of non-Gaussianity from inflation, say more than ... stars in the sky.
The most famous are:
\begin{itemize}
\item local NG, characteristic for multi-field, curvaton, ekpyrotic and cyclic models;
\item equilateral NG, associated to non-canonical kinetic and higher-derivative terms, DBI and K-inflation,  ghost inflation and EFT approaches;
\item orthogonal NG, which distinguishes between variants of non-canonical kinetic term and higher-derivative interactions;
\item flattened or folded NG.
\end{itemize}

\begin{figure}
  \centering
    \includegraphics[width=0.5\textwidth]{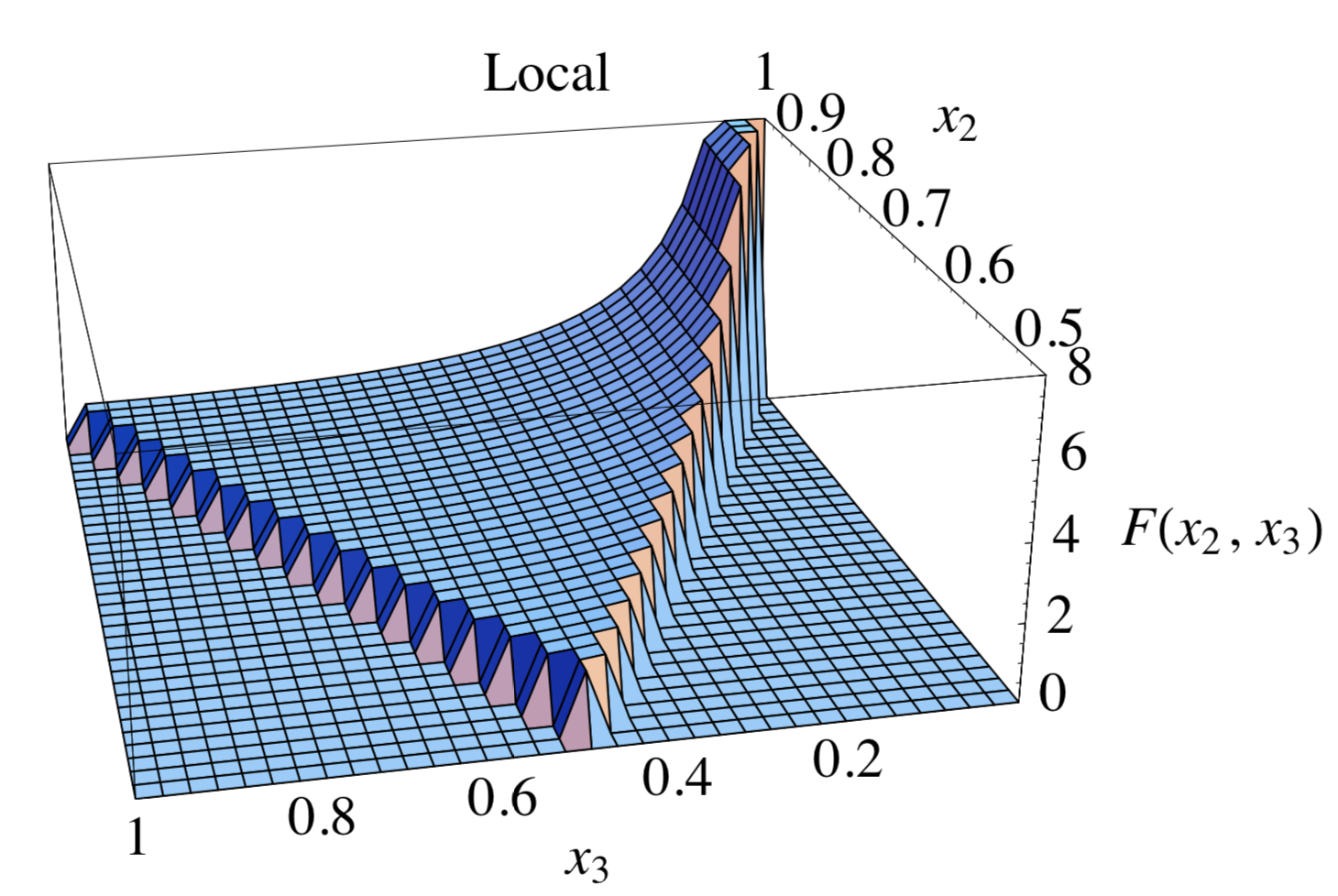}
    \caption{
    Plot of the function $F(1,x_2,x_3)x_2^2x^2_3$ for the local distribution. The figure is normalized to have value $1$ for equilateral configurations $x_2 = x_3 = 1$ and set to zero outside the region $1 - x_2 \le x_3 \le x_2$. From \cite{Babich:2004gb}.}
\label{LocalNG}
\end{figure}

\begin{figure} 
  \centering
    \includegraphics[width=0.5\textwidth]{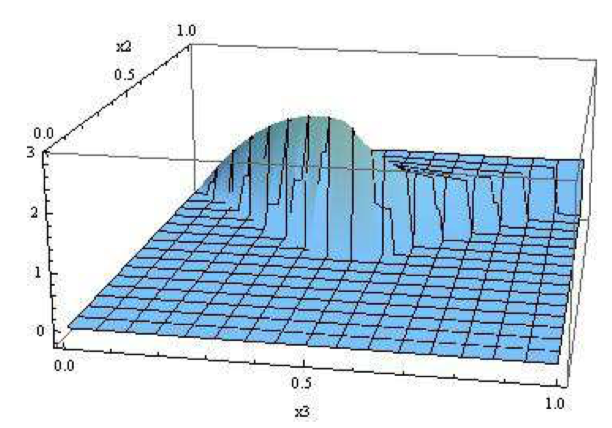}
    \caption{The folded template shape $F (x_2, x_3)  x_2^2x^2_3$, the maximum is  in the flat configuration ($k_1 = 1$, $k_2 = k_3=\frac{1}{2} $). From \cite{Bartolo:2010bj}.}
\label{FlattenedNG}
\end{figure}

More specifically, the  bispectrum for the local shape \cite{Salopek:1990jq, Gangui:1993tt, Verde:1999ij, Komatsu:2001rj}  peaks for squeezed triangles $k_3\ll k_1\sim k_2$, see Figure \ref{LocalNG}.
In this case, non-linearities develop outside the horizon during or immediately after inflation (e.g. multifield models of inflation).
\begin{figure}
  \centering
    \includegraphics[width=0.5\textwidth]{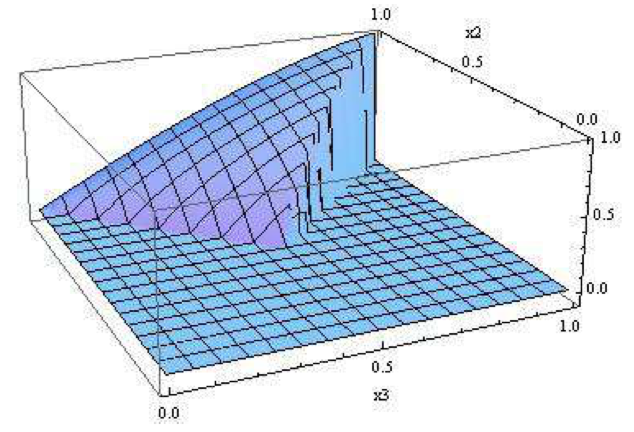}
    \caption{Plot of the function $F(1,x_2,x_3)x_2^2x^2_3$ for non-Gaussianities generated by higher derivative interactions. From \cite{Bartolo:2010bj}.}
\label{EquilateralNG}
\end{figure}
On the other hand, the bispectrum for the equilateral shape \cite{Creminelli:2005hu}, see Figure \ref{EquilateralNG}, peaks for  equilateral triangles $k_1=k_2=k_3$. 
Generally, in the equilateral family we can find single field models of inflation with non-canonical kinetic term $\mathcal{L}=P(\phi, X)$ with  $X=-\frac{1}{2}\partial_\mu \phi \, \partial^\mu \phi$ (e.g. DBI or K-inflation) where NG comes from higher derivative interactions  of the inflaton field, such as
\begin{equation}
\mathcal{L}\supset\delta\dot\phi (\nabla\delta\phi)^2 \, .
\end{equation} 
Finally, the bispectrum for the flattened shape  peaks for flattened (or folded) triangles $k_1 = k_2+k_3$, see Figure \ref{FlattenedNG}, and can be written in terms of the equilateral and orthogonal shapes  \cite{Senatore:2009gt}. It is characteristic for   excited initial states  (see \cite{Meerburg:2009ys, Chen:2006nt, Holman:2007na}),   higher derivative interactions  \cite{Bartolo:2010bj} or  models where a Galilean symmetry is imposed  \cite{Creminelli:2010qf}.

However, there are many other shapes: e.g. directionally dependent bispectra, tensor bispectra, etc.

\subsection{The role of $f_{\rm NL}$ and the detection of primordial non-Gaussianity}
\begin{figure}
    \includegraphics[width=0.45\textwidth]{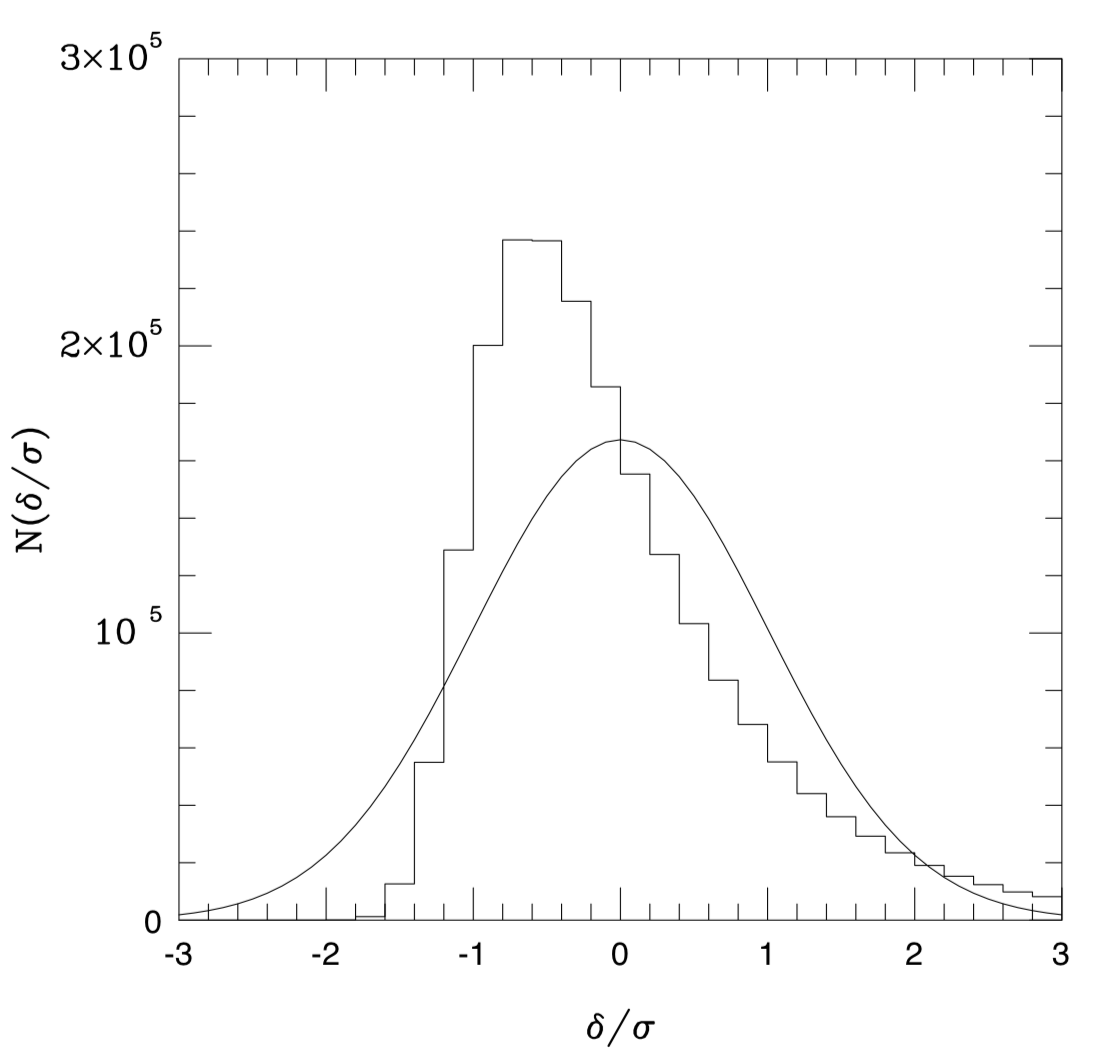}
     \includegraphics[width=0.45\textwidth]{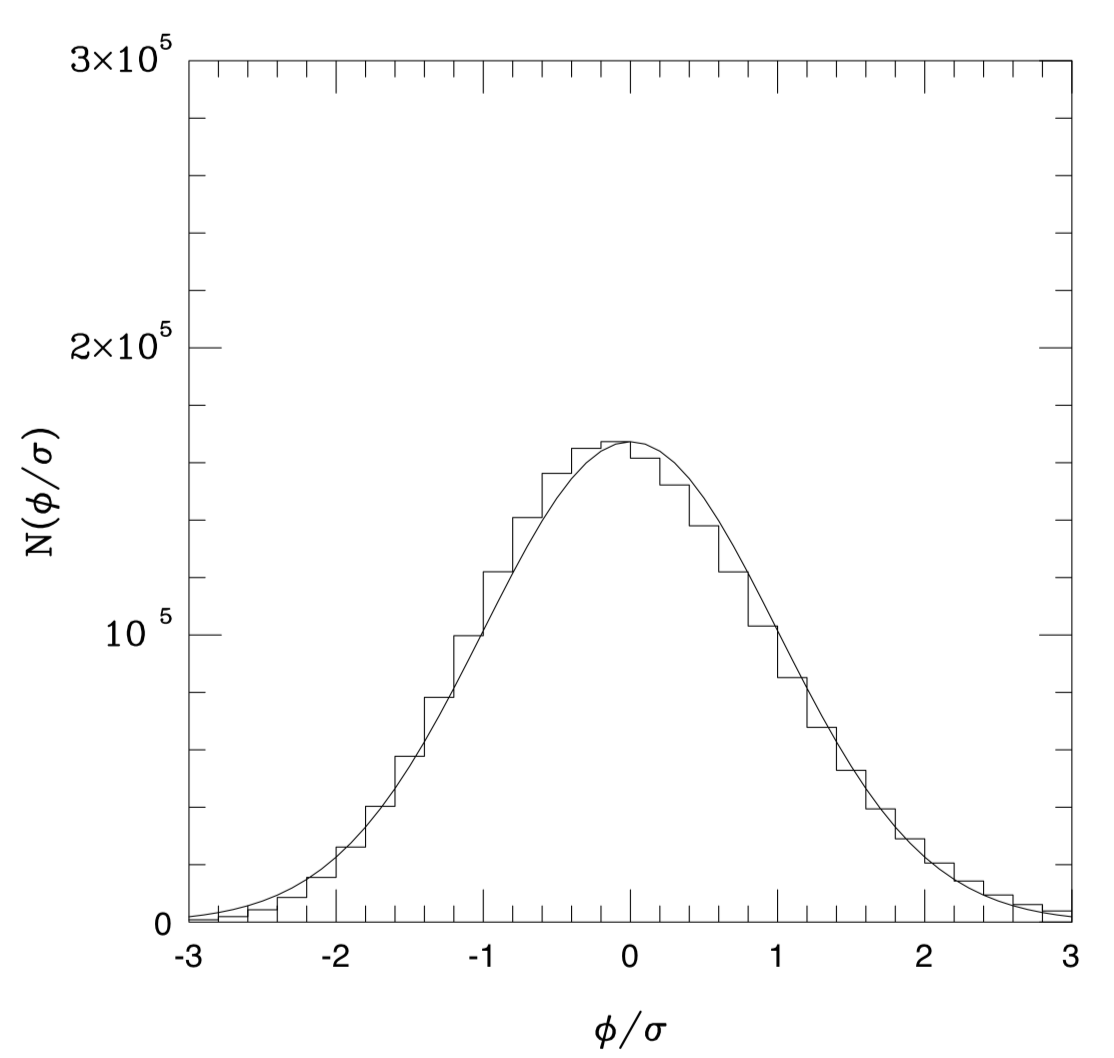}
    \caption{On the left, the distribution of densities. Solid curve is a Gaussian distribution with the same mean and variance. On the right, the distribution of potentials for the same model. Solid curve is a Gaussian distribution with the same mean and variance. From \cite{Scherrer:1994vg}.}
\label{DensityNonGaussian}
\end{figure}

Clearly, detecting a non-zero primordial bispectrum (e.g. $f_{\rm NL}\neq 0$) proves that the initial seeds were non-Gaussian. Similarly, for the trispectrum and  $n$-point correlation functions.
\\
However, the opposite is not true, namely detecting $f_{\rm NL}\approx 0$  doesn't prove Gaussianity.
Actually, there are infinitely many ways PNG can evade observational bounds optimized to search for $f_{\rm NL}$ and similar higher-order parameters.
\\
As an example, consider the situation where  the linear density contrast $\delta$ is non-Gaussian \cite{Scherrer:1994vg}.
In this case, by the central limit theorem, the gravitational potential $\Phi$ (which yields large-scale CMB anisotropies) tends to be much more Gaussian.
Indeed, consider the non-Gaussian distribution of densities  \cite{Scherrer:1994vg}
\begin{equation}
\delta(\mathbf{r})=\int f(|\mathbf{r}-\mathbf{r}'|)\Delta(\mathbf{r}')d^3\mathbf{r}'
\end{equation}
where $\Delta(\mathbf{r})$ is an uncorrelated field with a gamma distribution, and $f$ is chosen to give a Zel'dovich power spectrum ($P(k)\propto k$) for the density field. 
By solving Poisson's equation,  we get for the gravitational potential
\begin{equation}
\Phi(\mathbf{r})=-Ga^2\bar{\rho}\int \frac{\delta(\mathbf{r}')d^3\mathbf{r}'}{|\mathbf{r}-\mathbf{r}'|}
\end{equation}
with the resulting distribution very close to be Gaussian,  see Figure \ref{DensityNonGaussian}.
\section{Non-Gaussianity and Cosmic Microwave Background}
As mentioned before, the {\it Planck} satellite has provided accurate measurements of higher-order CMB correlations, resulting in  very stringent  constraints  on PNG. 
\\
{\it Planck} is a project of the European Space Agency, with instruments provided by two scientific Consortia funded by ESA member states (in particular the lead countries: France and Italy) with contributions from NASA (USA), and telescope reflectors provided in a collaboration between ESA and a scientific Consortium led and funded by Denmark.
\begin{figure}
  \centering
    \includegraphics[width=0.9\textwidth]{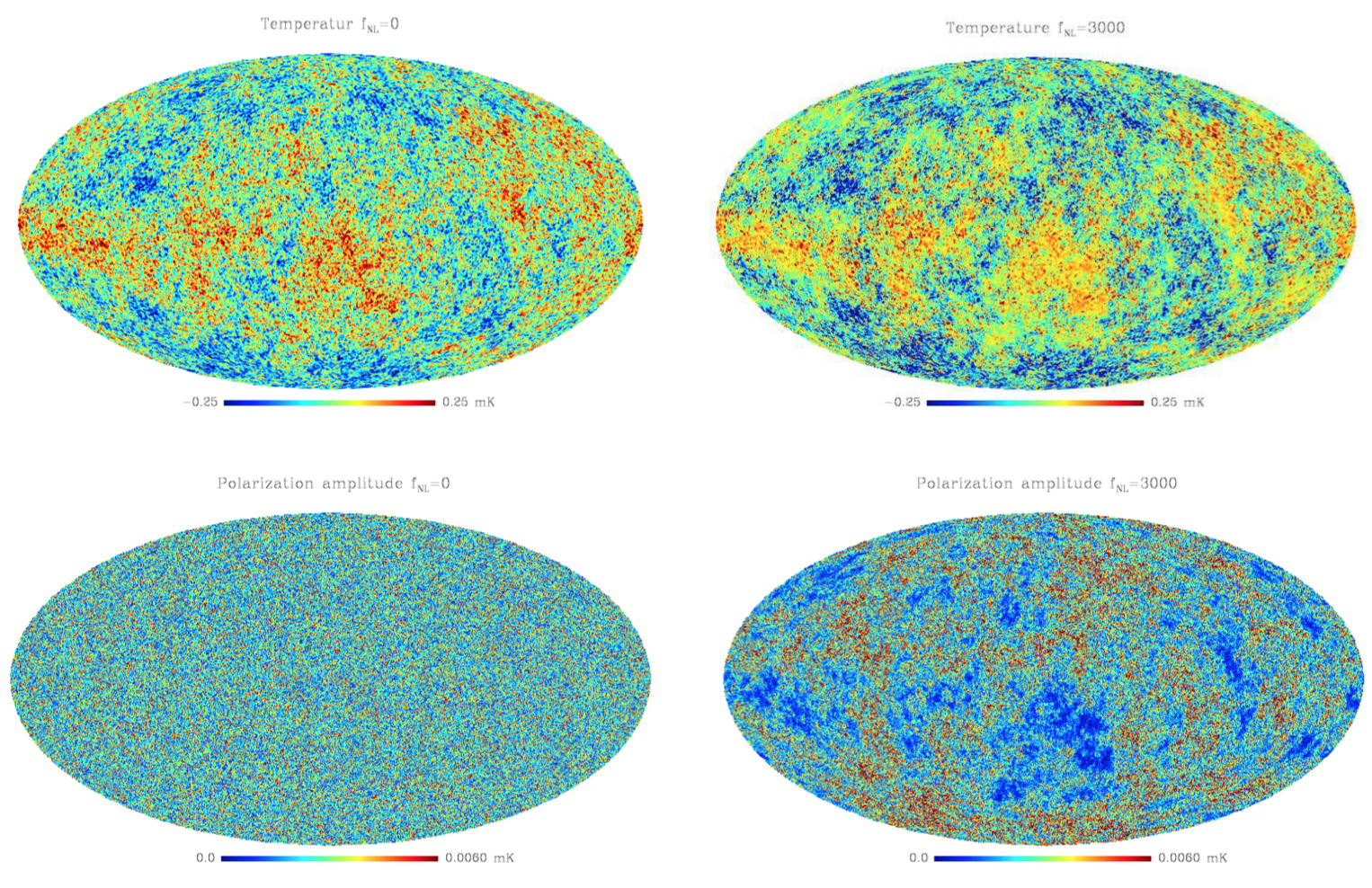}
    \caption{Left column: temperature and polarization intensity Gaussian CMB simulations. Right column: temperature and polarization non-Gaussian maps with the same Gaussian seed as in the left column and $f_{\rm NL} = 3000$. Temperatures are in $\text{mK}$. From \cite{Liguori:2007sj}.}
\label{NGPlanckMaps}
\end{figure}
The {\it Planck} satellite has  measured the CMB temperature and polarization  anisotropies with great accuracy, and, since PNG affects both, we can compare the data with the NG CMB simulated maps,  shown in Figure \ref{NGPlanckMaps}.

The latest release regarding non-Gaussianity  \cite{Ade:2015ava}  tested  the  local, equilateral, orthogonal (and many more) shapes  for the bispectrum and  provided new constraints on the primordial trispectrum parameter $g_{\rm NL}$ (while $\tau_{\rm NL}$ was constrained in the previous release \cite{Ade:2013ydc}).
A new {\it Planck} legacy release, which will improve the 2015 results in terms of more refined treatment of $E$-mode polarization,   is in preparation.

\begin{figure}
  \centering
    \includegraphics[width=0.5\textwidth]{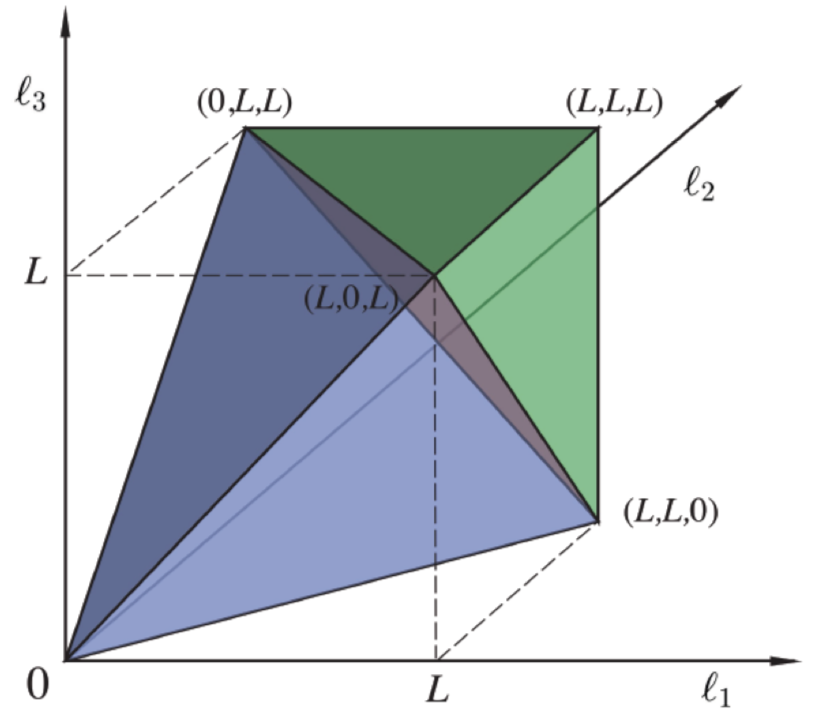}
    \caption{Permitted observational domain for the CMB bispectrum $b_{\ell_1\ell_2\ell_3}$. Allowed multipole values $(\ell_1, \ell_2, \ell_3)$ lie inside the shaded ``tetrapyd'' region (tetrahedron+pyramid), satisfying both the triangle condition and the experimental resolution. From \cite{Ade:2013ydc}.}
\label{NGAllowedRegion}
\end{figure}

The standard representation used in the {\it Planck} analysis for the CMB bispectrum is 
\begin{equation}
B^{m_1m_2m_3}_{\ell_1\ell_2\ell_3}\equiv \left<
a_{\ell_1m_1}a_{\ell_2m_2}a_{\ell_3m_3}\right>= \mathcal{G}_{m_1m_2m_3}^{\ell_1\ell_2\ell_3}b_{\ell_1\ell_2\ell_3}
\end{equation} 
where $\mathcal{G}_{m_1m_2m_3}^{\ell_1\ell_2\ell_3}$ are the Gaunt integrals
\begin{equation}
\mathcal{G}_{m_1m_2m_3}^{\ell_1\ell_2\ell_3}\equiv \int Y_{\ell_1m_1}(\hat{\mathbf{n}})Y_{\ell_2m_2}(\hat{\mathbf{n}})Y_{\ell_3m_3}(\hat{\mathbf{n}})d^2\hat{\mathbf{n}}
=
h_{\ell_1\ell_2\ell_3} \begin{pmatrix}
\ell_1 & \ell_2 & \ell_3 \\ 
m_1 & m_2 & m_3
\end{pmatrix} 
\end{equation}
and the $\ell_j$ satisfy the following conditions, see Figure \ref{NGAllowedRegion}:
\begin{itemize}
\item triangle  condition: $\ell_1\le \ell_2+\ell_3$ for $\ell_1\ge \ell_2, \ell_3$ and permutations
\item parity condition: $\ell_1+\ell_2+\ell_3=2 \ n$ with $n\in \mathbb{N}$
\item resolution: $ \ell_1, \ell_2, \ell_3\le \ell_{max}$ with $\ell_1, \ell_2, \ell_3\in \mathbb{N}$.
\end{itemize}
As noticed before, the search for PNG is optimized in terms of $f_{\rm NL}$.
Leaving aside complications coming from breaking of statistical isotropy 
(sky-cut, noise, etc.),   the general procedure is to  fit the theoretical bispectrum template
\begin{equation}
\hat{f}_{\rm NL}=\frac{1}{N} \sum B^{m_1m_2m_3}_{\ell_1\ell_2\ell_3} \left[   
\left(C^{-1}a\right)^{m_1}_{\ell_1}
\left(C^{-1}a\right)^{m_2}_{\ell_2}
\left(C^{-1}a\right)^{m_3}_{\ell_3}
-
3 C^{-1}_{\ell_1m_1\ell_2m_2}
\left(C^{-1}a\right)^{m_3}_{\ell_3}
\right]
\end{equation}
to  the 3-point  function obtained analyzing the data.
Unfortunately, a brute force implementation scales like  $  \ell_{max}^5  $, unfeasible at {\it Planck} 
(or WMAP) resolution.
On the other hand, we can  achieve a massive speed improvement ($\ell_{max}^3$   scaling)  if the reduced 
bispectrum is separable. 
Generally, there are different ways to write  the theoretical template  in separable form:
\begin{itemize}
\item  the  KSW \cite{Komatsu:2003iq} separable template fitting and the Skew-Cl  extension \cite{Munshi:2009ik};

\item  the binned bispectrum presented in \cite{Bucher:2009nm};

\item  the  modal expansion described in \cite{Fergusson:2009nv}.
\end{itemize}
The  alternative implementations differ basically in terms of the separation technique adopted and of the projection domain.

More recent improvements can be found  in \cite{Verde:2013gv}   where the interested reader can find an exact expression for the multi-variate joint probability distribution function (PDF) of non-Gaussian fields, primordially arising from local transformations of a Gaussian field.
This expression has been applied to the non-Gaussianity estimation from CMB maps and the halo mass function, obtaining  both analytical expressions as well as approximations with specified range of validity.
\\
The results  for the CMB  gave a fast way to compute the PDF, valid up to more than $7\, \sigma$ for $f_{\rm NL}$ values not ruled out by current observations, expressed as a combination of bispectrum and trispectrum of the temperature maps. 
Note that such expression is valid for any kind of non-Gaussianity and is not limited to the local type,  providing a useful basis for a fully Bayesian analysis of the NG parameter.

Finally, note that, in principle, we could go to higher order.
In fact, this may become important if we want to detect NG in observables characterized by a large $f_{\rm NL}$  (e.g. in high-redshift probes) and/or if  $f_{\rm NL}$ (leading order bispectrum) and $g_{\rm NL}$ (leading-order trispectrum) are both depending on the same underlying physical coupling constant that we aim at determining. 
However, the expressions are rather involved and we refer the reader to  \cite{Verde:2013gv} for details.

\subsection{{\it Planck} results on primordial non-Gaussianity}
In this section, we briefly present some of the {\it Planck} results on PNG, focusing on   the improvements compared to the 2013 release and  the   differences between the  methods used in the analysis, particularly concerning  the  Integrated Sachs-Wolfe (ISW) effect.

\begin{figure}
  \centering
    \includegraphics[width=0.8\textwidth]{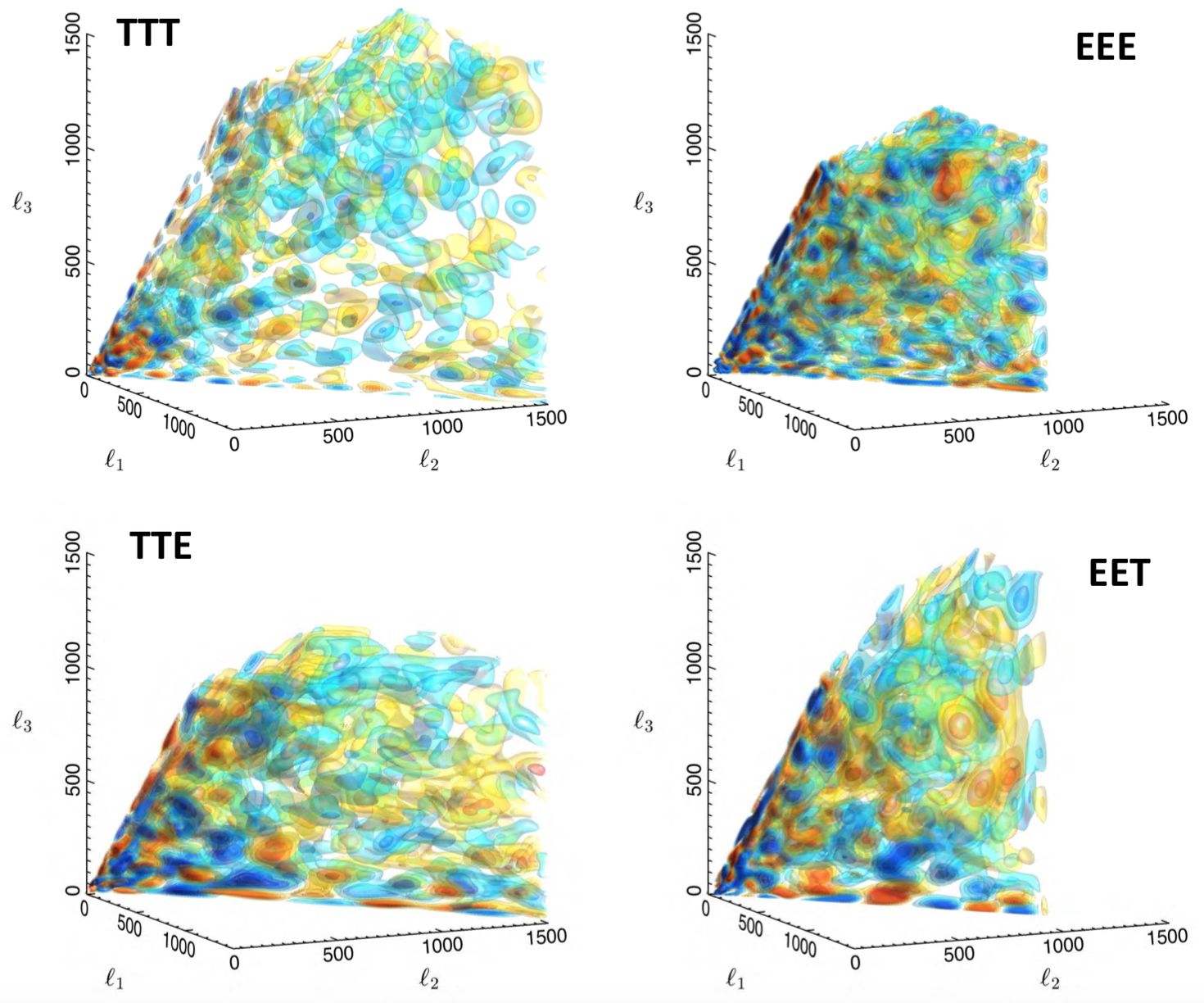}
    \caption{ CMB temperature and polarization bispectrum reconstructions for {\it Planck} \SMICA\ maps using the full set of polynomial modes with $n_{max} = 2001$ and with signal-to-noise weighting. From \cite{Ade:2015ava}.}
\label{PlanckNGTTTEEE}
\end{figure}

Let us start with the 2015 {\it Planck} analysis for the bispectrum in the modal representation,  described in Figure \ref{PlanckNGTTTEEE}, for the various combinations  $TTT$, $EEE$, $TTE$, $EET$.
\\
Note that compared to {\it Planck} 2013 the new constraints on local, equilateral, orthogonal bispectra have improved by up to 15\%.

\begin{figure}
\centering
\includegraphics[width=0.7\textwidth]{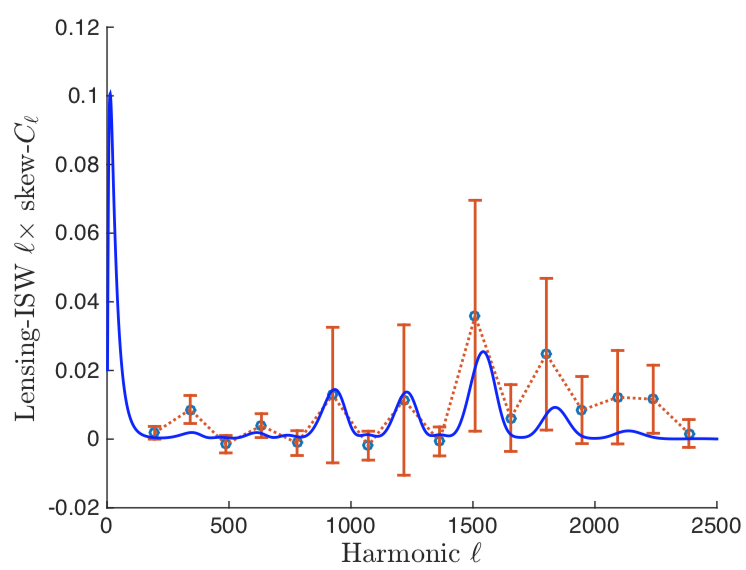}
\caption{The skew-Cl spectrum for the lensing-ISW effect (red line with data points), from the temperature map. The blue curve is the theoretically-expected spectrum. From \cite{Ade:2015ava}. }
\label{Lensing_ISW}
\end{figure}

In the 2015 analysis particular attention was devoted to investigate  the ISW-lensing effect which strongly affects the constraints on $f_{\rm NL}$ from {\it Planck} bispectrum. 
The results are confirmed using several $f_{\rm NL}$  estimators and  CMB maps. Actually, different maps have been produced by the \SMICA, \NILC, \SEVEM \hspace{1mm} and \Commander -\texttt{Ruler}  (or \texttt{C-R}) pipelines. Notice that the \SMICA\ product is considered the preferred one overall. 
Having said this, the amplitude of the ISW-lensing bispectrum from the \SMICA, \NILC, \SEVEM, and \texttt{C-R}\   foreground-cleaned maps, for the KSW, binned, and modal (polynomial) estimators  are summarized in Table \ref{table:fNL_lisw}.
Remarkably, the coupling between weak lensing and  ISW effect is the leading contamination to local NG, as the ISW lensing bispectrum has been detected  with a significance of $2.8 \, \sigma$ (see Figure \ref{Lensing_ISW}), and improves to $3.0 \, \sigma$ when including polarization. 
In conclusion, the bias in the three primordial $f_{\rm NL}$ parameters due to the ISW-lensing signal is described in Table \ref{tab:lisw_bias}.

\begin{table}[ht]               
\begingroup
\footnotesize
\setbox\tablebox=\vbox{
   \newdimen\digitwidth
   \setbox0=\hbox{\rm 0}
   \digitwidth=\wd0
   \catcode`*=\active
   \def*{\kern\digitwidth}
   \newdimen\signwidth
   \setbox0=\hbox{+}
   \signwidth=\wd0
   \catcode`!=\active
   \def!{\kern\signwidth}
\halign{\hbox to 0.65in{#\leaderfil}\tabskip 1em&
\hfil#\hfil\tabskip 0.7em&
\hfil#\hfil\tabskip 0.7em&
\hfil#\hfil\tabskip 0.7em&
\hfil#\hfil\tabskip 0pt\cr
\noalign{\doubleline\vskip 2pt}
\omit&\multispan4\hfil lensing-ISW amplitude\hfil\cr
\omit&\multispan4\hrulefill\cr
Method\hfill&\SMICA&\SEVEM&\NILC&\Commander\cr
\noalign{\vskip 4pt\hrule\vskip 6pt}
\omit\hfil \itT\hfil&&\cr
KSW & 0.79 $\pm$ 0.28 & 0.78 $\pm$ 0.28 & 0.78 $\pm$ 0.28 & 0.84 $\pm$ 0.28 \cr
Binned & 0.59 $\pm$ 0.33 & 0.60 $\pm$ 0.33 & 0.68 $\pm$ 0.33 & 0.65 $\pm$ 0.36 \cr
Modal2 & 0.72 $\pm$ 0.26 & 0.73 $\pm$ 0.26 & 0.73 $\pm$ 0.26 & 0.78 $\pm$ 0.27 \cr
\noalign{\vskip 4pt\hrule\vskip 6pt}
\omit\hfil \textit{T+E}\hfil&&\cr
Binned & 0.82 $\pm$ 0.27 & 0.75 $\pm$ 0.28 & 0.85 $\pm$ 0.26 & 0.84 $\pm$ 0.27 \cr
\noalign{\vskip 3pt\hrule\vskip 4pt}}}
\endPlancktable                 
\endgroup
\caption{Results for the amplitude of the lensing-ISW bispectrum from the
\SMICA, \SEVEM, \NILC, and \Commander\ foreground-cleaned maps,
for different bispectrum estimators. Error bars are 68\,\%~CL. From \cite{Ade:2015ava}.
}
\label{table:fNL_lisw}   
\end{table}                     

Actually, the new analysis was the first  adopting also polarization data for PNG, and the {\it Planck} 2015 constraints on $T+E$ have confirmed $T$ results with significantly reduced error bars, see Table  \ref{Tab_KSW+SMICA} for the latest constraints on $f_{\rm NL}$ for the various shapes.
\\
Specifically, the {\it Planck} 2013 hints of NG in oscillatory feature models remain in $T$, but decrease significantly when polarization is included. 
 Also,  new estimators for high-frequency oscillations cover 10 times more parameter space, compared to the previous analysis.
 
\begin{table}[ht]            
\begingroup
\setbox\tablebox=\vbox{
   \newdimen\digitwidth
   \setbox0=\hbox{\rm 0}
   \digitwidth=\wd0
   \catcode`*=\active
   \def*{\kern\digitwidth}

   \newdimen\signwidth
   \setbox0=\hbox{+}
   \signwidth=\wd0
   \catcode`!=\active
   \def!{\kern\signwidth}

   \newdimen\dotwidth
   \setbox0=\hbox{.}
   \dotwidth=\wd0
   \catcode`^=\active
   \def^{\kern\dotwidth}

\halign{
\hbox to 1.1in{#\leaderfil}\tabskip 1em&
\hfil#\hfil&
\hfil#\hfil&
\hfil#\hfil&
\hfil#\hfil\tabskip 0pt\cr
\noalign{\doubleline\vskip 2pt}
\omit&\multispan4\hfil lensing-ISW $f_{\rm NL}$ bias\hfil\cr
\omit&\multispan4\hrulefill\cr
Shape\hfill&\SMICA&\SEVEM&\NILC&\Commander\cr
\noalign{\vskip 4pt\hrule\vskip 6pt}
\itT\ Local & !*7.5& !*7.5& !*7.3& !*7.0 \cr
\itT\ Equilateral & !*1.1& !*1.2& !*1.3& !*1.8 \cr
\itT\ Orthogonal & $-$27^*& $-$27^*& $-$26^*& $-$26^*\cr
\noalign{\vskip 4pt\hrule\vskip 6pt}
\itE\ Local & !*1.0& !*1.1& !*1.0& !*1.1 \cr
\itE\ Equilateral & !*2.6& !*2.7& !*2.5& !*2.9 \cr
\itE\ Orthogonal & !$-$1.3& !$-$1.3& !$-$1.2& !$-$1.5 \cr
\noalign{\vskip 4pt\hrule\vskip 6pt}
\itTpE\ Local & !*5.2& !*5.5& !*5.1& !*4.9\cr
\itTpE\ Equilateral & !*3.4& !*3.4& !*3.4& !*3.6\cr
\itTpE\ Orthogonal & $-$10^*& $-$11^*& $-$10^*&$-$10^* \cr
\noalign{\vskip 3pt\hrule\vskip 4pt}}}
\endPlancktable                    

\endgroup
\caption{Bias in the three primordial $f_{\rm NL}$ parameters due to the
lensing-ISW signal for the four component separation methods. From \cite{Ade:2015ava}.}
\label{tab:lisw_bias}  
\end{table}                      

\begin{table}[ht]              
\begingroup
\setbox\tablebox=\vbox{
   \newdimen\digitwidth
   \setbox0=\hbox{\rm 0}
   \digitwidth=\wd0
   \catcode`*=\active
   \def*{\kern\digitwidth}
   \newdimen\signwidth
   \setbox0=\hbox{+}
   \signwidth=\wd0
   \catcode`!=\active
   \def!{\kern\signwidth}
\newdimen\dotwidth
\setbox0=\hbox{.}
\dotwidth=\wd0
\catcode`^=\active
\def^{\kern\dotwidth}
\halign{\hbox to 1in{#\leaderfil}\tabskip 1em&
\hfil#\hfil\tabskip 1em&
\hfil#\hfil\tabskip 0pt\cr
\noalign{\vskip 10pt\doubleline\vskip 2pt}
\omit&\multispan2\hfil $f_{\rm NL}$(KSW)\hfil\cr
\omit&\multispan2\hrulefill\cr
Shape and method\hfill&\hfil Independent\hfil&
\hfil ISW-lensing subtracted\hfil\cr
\noalign{\vskip 2pt}
\noalign{\vskip 4pt\hrule\vskip 6pt}
\omit\hfil \SMICA\,\, (\itT) \hfil&\cr
Local& !10.2!$\pm$!*5.7&
{\bf*!2.5!$\pm$!*5.7}\cr
Equilateral& $-$13*^!$\pm$!70^*&
{\bf$-$16*^!$\pm$!70^*}\cr
Orthogonal& $-$56*^!$\pm$!33^*&
{\bf$-$34^*!$\pm$!33^*}\cr
\noalign{\vskip 5pt}
\omit\hfil {\SMICA\,\,(\itTpE) }\hfil&\cr
Local& *!6.5!$\pm$!*5.0&
{\bf*!0.8!$\pm$!*5.0}\cr
Equilateral& *!3^*!$\pm$!43^*&
{\bf*$-$4^*!$\pm$!43^*}\cr
Orthogonal& $-$36^*!$\pm$!21^*&
{\bf$-$26^*!$\pm$!21^*}\cr
\noalign{\vskip 3pt\hrule\vskip 4pt}}}
\endPlancktable                   
\endgroup
\caption{Results for the $f_{\rm NL}$ parameters of the primordial local, 
equilateral, and orthogonal shapes, determined by the KSW estimator from the
\SMICA\ foreground-cleaned map. 
 Error bars are $68\,\%$ CL. 
 From \cite{Ade:2015ava}.}
\label{Tab_KSW+SMICA}
\end{table}                        

Remarkably, the improvements in the {\it Planck} 2015 results have allowed to put new constraints on: 
\begin{itemize}
\item isocurvature NG, where polarizartion data were crucial in this respect;
\item tensor NG, where parity-odd $T$ limits are consistent with WMAP (null result);
\item  trispectrum due to cubic NG (in particular $g_{\rm NL}$ for a variety of shapes).
\end{itemize}
Finally, with  the 2015 release  we could also  constraint the three fundamental shapes of the trispectrum 
\begin{equation}
\begin{aligned}
g_{\rm NL}^{local}&=(-9.0\pm 7.7)\times 10^4 \, ,
\\
g_{\rm NL}^{\dot{\sigma}^4}&=(-0.2\pm 1.7)\times 10^6 \, , 
\\
g_{\rm NL}^{(\partial{\sigma})^4}&=(-0.1\pm 3.8)\times 10^5 \, .
\end{aligned}
\end{equation}
In conclusion, the  {\it Planck}  2015 release contains a  largely extended analysis of NG templates, and we expect that the
upcoming release (the ``{\it Planck} legacy'' paper) will further improve the constraints on standard shapes (owing to refined treatment of $E$-mode polarization maps) and add some extra shapes (such as scale-dependent $f_{\rm NL}$, conformal symmetry),  looking for features  both in the power spectrum and the bispectrum.
\subsection{Implications for inflation}
One of the most important consequence from {\it Planck} data is that the  simplest inflationary models (standard inflation) are still alive  ... and in very good shape!
\\
Specifically, for standard inflation we refer to a  single scalar field $\phi$ (representing a single clock), characterized by a Bunch-Davies initial vacuum state and a  canonical kinetic term $X=-\frac{1}{2}\partial_\mu \phi \, \partial^\mu \phi$, performing a slow-roll dynamics by means of a potential $\mathcal{V}(\phi)$,  minimally coupled to  gravity, described  by general relativity (GR)
\begin{equation}
S=\int d^4x\sqrt{-g}\left[ X-\mathcal{V}(\phi)+\mathcal{L}_{int}(\phi,A_\mu,\Psi)+\frac{M_{Pl}^2}{2} \, 
R
\right]
\end{equation}
where $\mathcal{L}_{int}$ is the interaction term between the inflaton  and  other fields  such as gauge bosons $A_\mu$ or fermions $\Psi$. 
\\
Actually, standard inflation predicts tiny  ($\mathcal{O}(10^{-2})$, thus no presently detectable)   PNG.

However,  alternatives to standard inflation have also been considered. In particular,  results from  {\it Planck} 2015 (which increased the number of modes  from 600 to 2000 with respect to {\it Planck} 2013)  constrained $f_{\rm NL}$  for a large number of inflationary models  including:
\begin{itemize}
\item the equilateral family (DBI, EFT, ghost and K-inflation);
\item the flattened shapes (non-Bunch Davies);
\item feature models (oscillatory or  scale-dependent bispectra);
\item direction dependence;
\item quasi-single-field;
\item parity-odd models.
\end{itemize}
Since no evidence for NG has been found,   we could only  put tighter constraints on the parameters from the models above,  for example on:
\begin{itemize}
\item the  curvaton decay fraction $r_D > 19\%$ (from local $f_{\rm NL}$, $T+E$),
\item the speed of sound in the Effective Field Theory of Inflation \cite{Cheung:2007st} $c_S > 0.024$ (from equilateral and  orthogonal $f_{\rm NL}$), 
\item the speed of sound in DBI inflation $c_S > 0.087$ (from $T+E$).
\end{itemize}

\subsection{Primordial non-Gaussianity with CMB spectral distorsions}
Possible measurements to improve the constraints on PNG include  CMB spectral distorsions from acoustic wave dissipation that can  probe a large range  of scales, much more than CMB/LSS \cite{Khatri:2013dha}. 

Actually,  if $\mu$-anisotropies were measured we would access to 
\begin{itemize}
\item $T\mu$ correlations,  useful to investigate the primordial local $f_{\rm NL}$ (see \cite{Pajer:2013oca}) or    other squeezed shapes, e.g. excited initial states \cite{Ganc:2012ae},
\item  $\mu\mu$ correlations, associated to the primordial local trispectrum, $\tau_{\rm NL}$ \cite{Bartolo:2015fqz},

\item $TT\mu$ bispectrum, related to the primordial local trispectrum $g_{\rm NL}$ \cite{Bartolo:2015fqz}.
\end{itemize}
While for Gaussian initial conditions the dissipated power  in small patches (from
$\approx 50 \ \text{Mpc}^{-1}$ to $\approx 10^{4}\  \text{Mpc}^{-1}$) is isotropically distributed, squeezed bispectra associated to local NG generate couplings between large and small scales.
\\
As a consequence of these coupling between long and short modes, the CMB temperature fluctuations on large scales can be coupled to spectral distortions arising from acoustic wave dissipation at very small scales, resulting in $T\mu$ correlations
\cite{Pajer:2012vz, Pajer:2013oca, Emami:2015xqa}. 
\\
More specifically, following \cite{Emami:2015xqa}, consider the curvature perturbation at position $\vec{x}$ in terms of a Gaussian random variable $z(\vec{x})$ 
\begin{equation}
\zeta(\vec{x})=z(\vec{x})+ \frac{3}{5}f^{loc}_{\rm NL}z^2(\vec{x}) \, . 
\end{equation}
Splitting $\zeta(\vec{x})$ in  long-wavelength  and short-wavelength modes  as $\zeta(\vec{x}) =\zeta_L(\vec{x})+\zeta_S(\vec{x})$, and similarly writing $z(\vec{x})=z_L(\vec{x})+z_S(\vec{x})$ we get 
\begin{equation}
\zeta_L+\zeta_S=z_L+z_S+\frac{3}{5}f^{loc}_{\rm NL}[z_L^2+2 \, z_L\, z_S+z_S^2] \, ,
\end{equation}
and we conclude that,  in presence of  $f_{\rm NL}^{loc}\neq 0$,  long and short modes can be coupled.  
In particular, to linear order in $f_{\rm NL}^{loc}$,  the small-scale curvature fluctuation in the presence of some fixed long-wavelength curvature fluctuation is
\begin{equation}
\zeta_S=z_S\left(1+\frac{6}{5}f^{loc}_{\rm NL} \zeta_L\right) 
\end{equation}
thus modulating  the (fractional) chemical-potential fluctuation,  given by \cite{Pajer:2012vz, Emami:2015xqa, Cabass:2017kho} 
\begin{equation}
\frac{\delta\mu}{\mu}\approx \frac{\delta\left< \zeta^2\right>}{\left< \zeta^2\right>}\approx \frac{12}{5}f^{loc}_{\rm NL} \zeta_L \, .
\end{equation}
Similarly, for  the  large-angle ($\ell \lesssim  100$, probing causally disconnected regions at the
  last scattering surface) the temperature fluctuation is  determined primarily by the curvature fluctuation at the surface of last scatter, given by $\delta T/T \approx \zeta/5$ \cite{Pajer:2012vz, Emami:2015xqa}.
\\
As a consequence,  the fractional
chemical potential fluctuation $\frac{\delta\mu}{\mu}$ and the temperature fluctuation $\frac{\delta T}{T}$  are cross correlated with an angular power spectrum $C_\ell^{\mu T}$  equal to  \cite{Pajer:2012vz, Emami:2015xqa}
\begin{equation}
C_\ell^{\mu T}=12 f^{loc}_{\rm NL} C_\ell^{TT} \, .
\end{equation}

\section{Primordial non-Gaussianity and the Large-Scale Structure}
Non-Gaussianity in Large-Scale Structure can be either of primordial origin or associated to gravitational instability.
\\
In particular, to make contact with the CMB definition, PNG in LSS  can be defined  starting from the DM density fluctuation $\delta$ through the  Poisson's
equation 
\begin{equation}
\delta=-\left( \frac{3}{2}\Omega_m H^2
\right)^{-1}\nabla^2\Phi
\end{equation}
where we have used the  comoving gauge  for density fluctuation \cite{Bardeen:1980kt}. 
\\
As before, we can write
\begin{equation}
\Phi=\phi_L+f_{\rm NL}(\phi_L^2-\left<
\phi_L^2\right>)
+g_{\rm NL}(\phi^3-\left< \phi_L^2\right>\phi_L)+\dots
\end{equation}
where $\phi_L$ is the linear Gaussian contribution and $f_{\rm NL}$ and $g_{\rm NL}$ are dimensionless non-linearity parameters
\footnote{CMB and LSS conventions may differ by a factor $1.3$ for $f_{\rm NL}$, $(1.3)^2$ for $g_{\rm NL}$.}.

\begin{figure}
  \centering
    \includegraphics[width=0.9\textwidth]{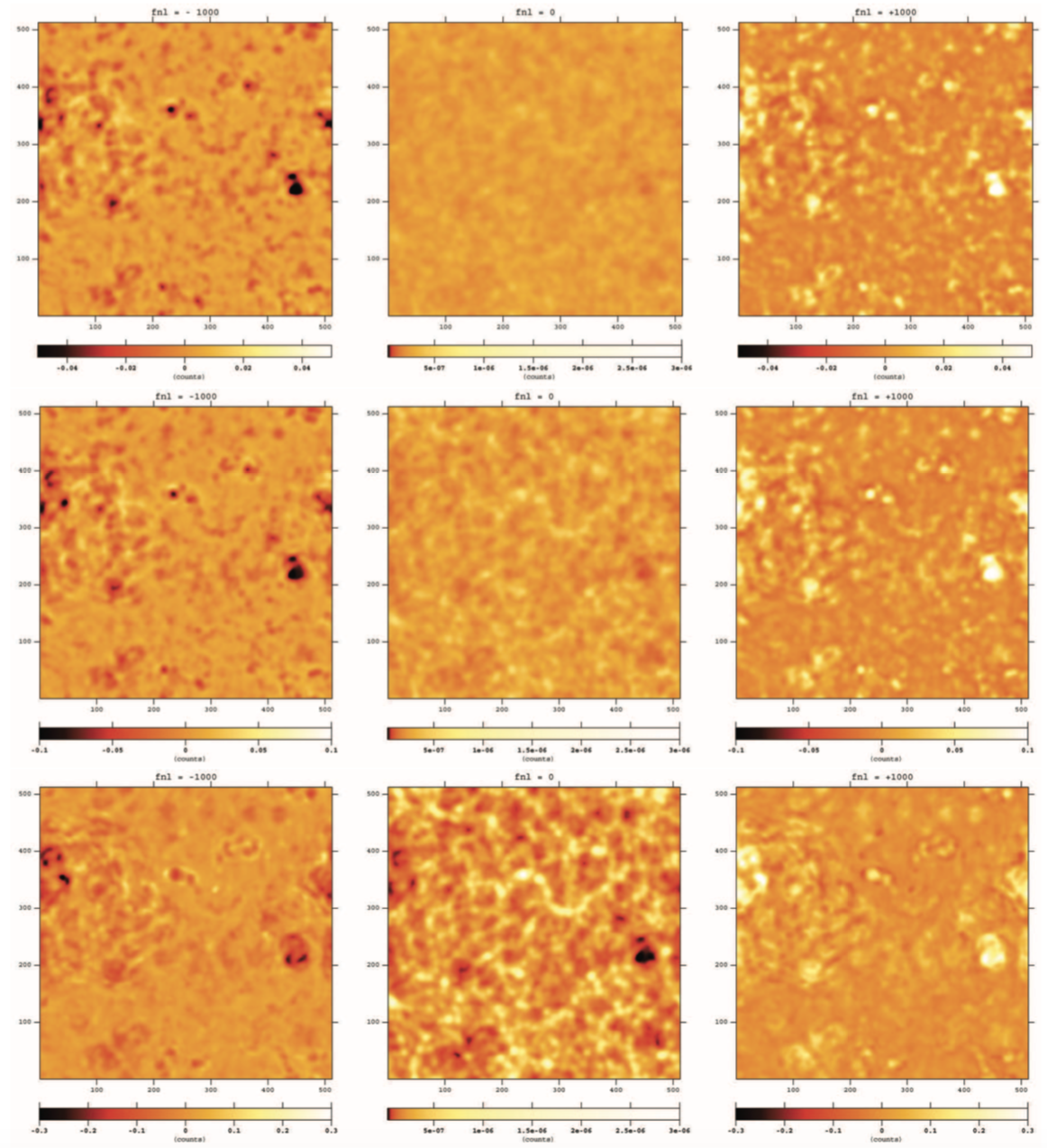}
    \caption{Slice maps of simulated mass density fields at $z = 5.15$ (top), $z = 2.13$ (middle) and $z = 0$ (bottom). The number of pixels at a side length is $512$ ($500h^{-1}$ Mpc) and that of the thickness is $32$ ($31.25h^{-1}$ Mpc). The panels in the middle row show the log of the projected density smoothed with a Gaussian filter of $10$ pixels width, corresponding to $9.8h^{-1}$ Mpc. The left and right panels are the relative residuals for the $f_{\rm NL}=\pm 1000$ runs. Each panel has the corresponding color bar and the range considered are different from panel to panel. From \cite{Hikage:2007bc}.}
\label{Hikage}
\end{figure}

In order to investigate PNG in LSS, generally N-body simulations have been playing a crucial role \cite{Grossi:2007ry, Hikage:2007bc, Desjacques:2008vf, Giannantonio:2009ak, Grossi:2009an, Sefusatti:2010ee, Wagner:2010me, Wagner:2011wx}. The standard equations  are
\begin{equation}
\begin{aligned}
\Phi &=\phi_L+f_{\rm NL}(\phi_L^2-\left< \phi_L^2\right>)
\\
\nabla^2(\Phi \ast T)g(z)&=-4\pi Ga^2 \delta\rho_{DM}
\end{aligned}
\end{equation}
where $T$ is the  matter transfer function and $g$ is the growth suppression factor. Typical results are shown in Figure \ref{Hikage}.

\begin{figure}
      \includegraphics[width=0.9\textwidth]{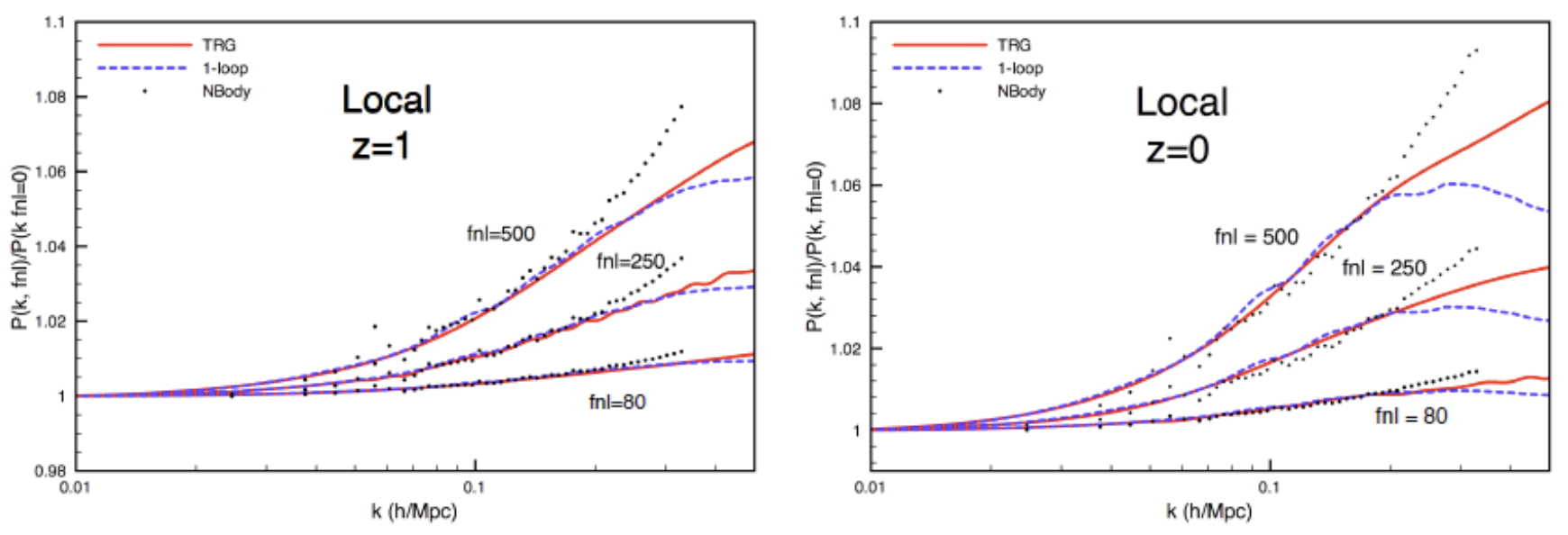}
    \caption{Ratio of the non-Gaussian to Gaussian power spectrum for several values of $f_{\rm NL}$ in the local model. The dots correspond to the data from the N-body simulations of \cite{Pillepich:2008ka}. The red (continuous) line is the TRG result of this paper and the blue (dashed) line is the one-loop result. From \cite{Bartolo:2009rb}.}
\label{figures_Ratios_fnl_Ratio_Local_z1}
\end{figure}

\begin{figure}
 \includegraphics[width=0.9\textwidth]{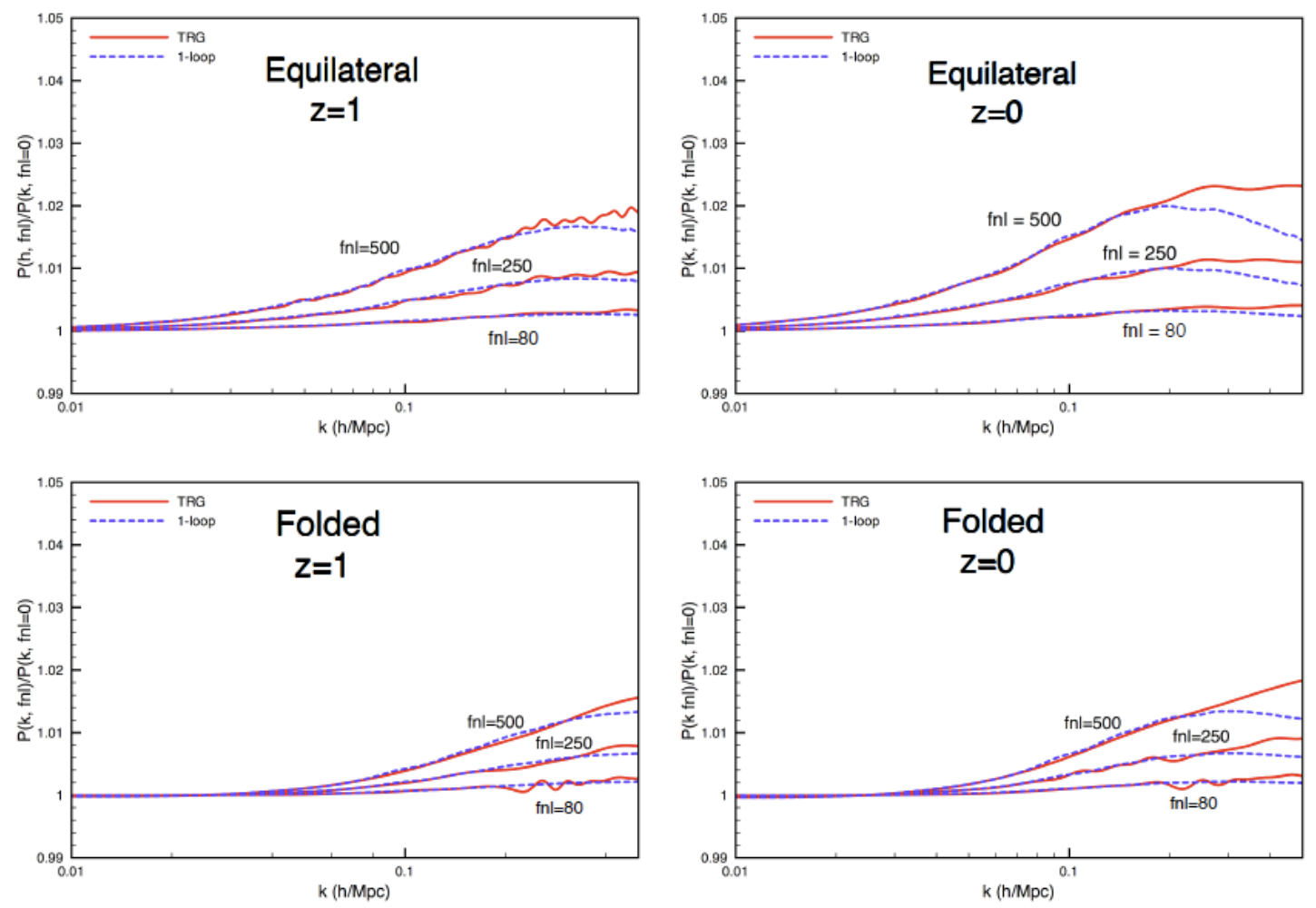}
    \caption{Ratio of the non-Gaussian to Gaussian power spectrum for several values of $f_{\rm NL}$ in the equilateral (top panels) and folded (bottom panels) models. The red (continuous) lines are the TRG result of this paper and the blue (dashed) lines are the one-loop result. From \cite{Bartolo:2009rb}.}
\label{figures_Ratios_fnl_Ratio_EQFS_z1}
\end{figure}

However, in the mild non-linear regime, analytical approaches have been developed to study the PNG effects on the matter power spectrum. Results for the local shape calculated using the Time Renormalization Group theory   \cite{Matarrese:2007wc, Pietroni:2008jx} compared to N-body are shown in Figure \ref{figures_Ratios_fnl_Ratio_Local_z1}, while for the equilateral and folded shapes see Figure  \ref{figures_Ratios_fnl_Ratio_EQFS_z1}.
\\
 Similar techniques include the renormalized perturbation theory \cite{Crocce:2005xy, Bernardeau:2008fa, Montesano:2010qc}, renormalization group approach \cite{McDonald:2006hf}, closure theory \cite{Taruya:2007xy}, Lagrangian perturbation theory \cite{Buchert:1992ya, Matsubara:2008wx, Zheligovsky:2013eca}, the  time-sliced perturbation theory \cite{Blas:2015qsi} and the Effective Field Theory of LSS (EFTofLSS) \cite{Baumann:2010tm, Carrasco:2012cv}.
 Specifically, the EFTofLSS for non-Gaussian initial conditions  have been developed in  \cite{Assassi:2015jqa}, see also \cite{Hertzberg:2012qn, Carrasco:2013mua, Porto:2013qua, Senatore:2014vja, Baldauf:2014qfa, Angulo:2015eqa, Abolhasani:2015mra}.

\subsection{Non-Gaussianity and halo mass function}
Besides using the standard statistical estimators, like the (mass) bispectrum, trispectrum, etc., one can look at the tails of the distribution, i.e. at rare events. 

Rare events have the advantage that they often maximize deviations from what is predicted by a Gaussian distribution, but have the obvious disadvantage of being rare! But remember that, according to the Press-Schechter-like schemes, all collapsed DM halos correspond to (rare) high peaks of the underlying density field.

\begin{figure}
  \centering
    \includegraphics[width=0.8\textwidth]{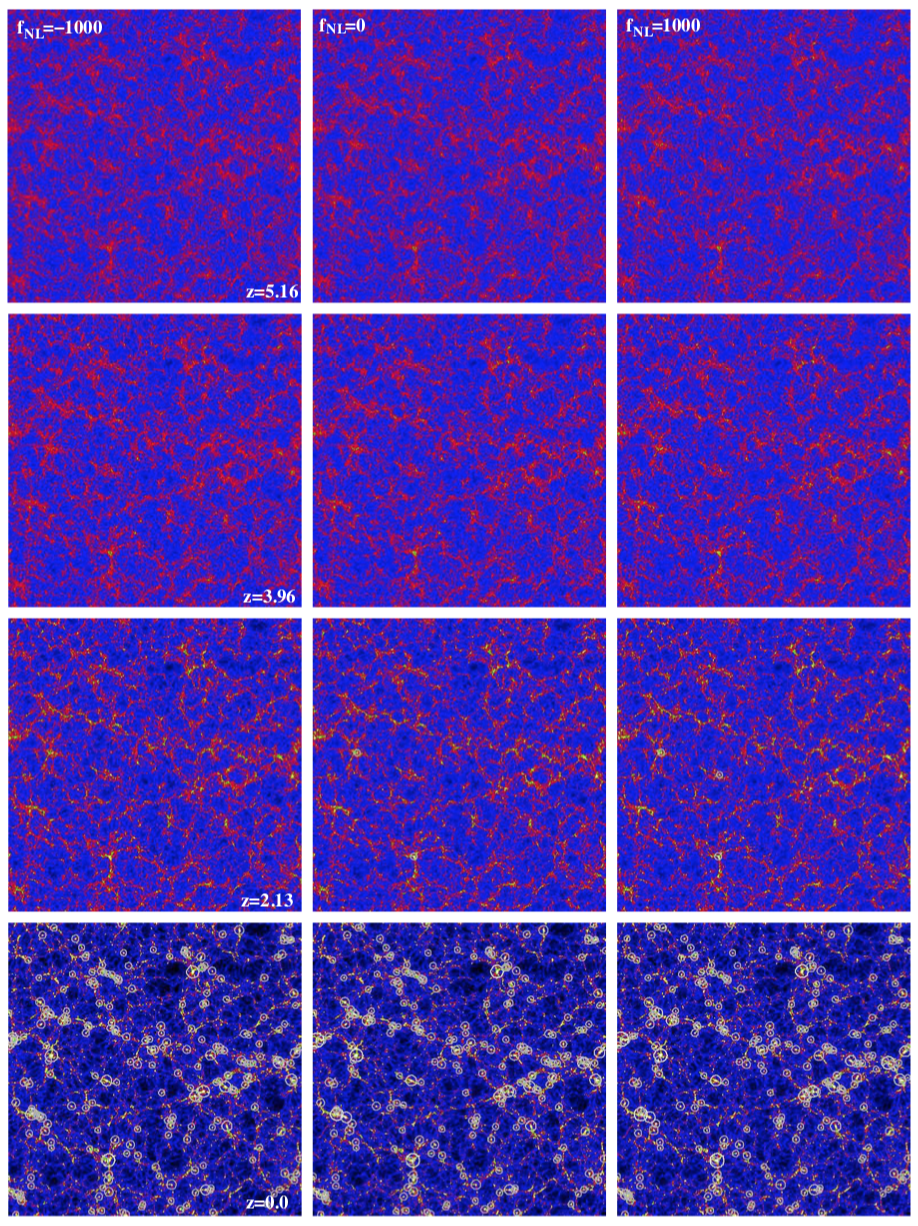}
    \caption{Mass density distribution and halo positions in a slice cut across the simulation box. The color-coded contours indicate different density levels ranging from dark (deep blue) underdense regions to bright (yellow) high density peaks. The halo positions are indicated by open circles with size proportional to their masses. Left panels: NG model with $f_{\rm NL} = -1000$. Central panels: Gaussian model. Right panels NG model with $f_{\rm NL} = +1000$. The mass and halo distributions are shown at various epochs, characterized by increasing redshifts (from bottom to top), as indicated in the panels. From \cite{Grossi:2007ry}.}
\label{GrossiSimulationNG}
\end{figure}

\begin{figure}
  \centering
    \includegraphics[width=0.9\textwidth]{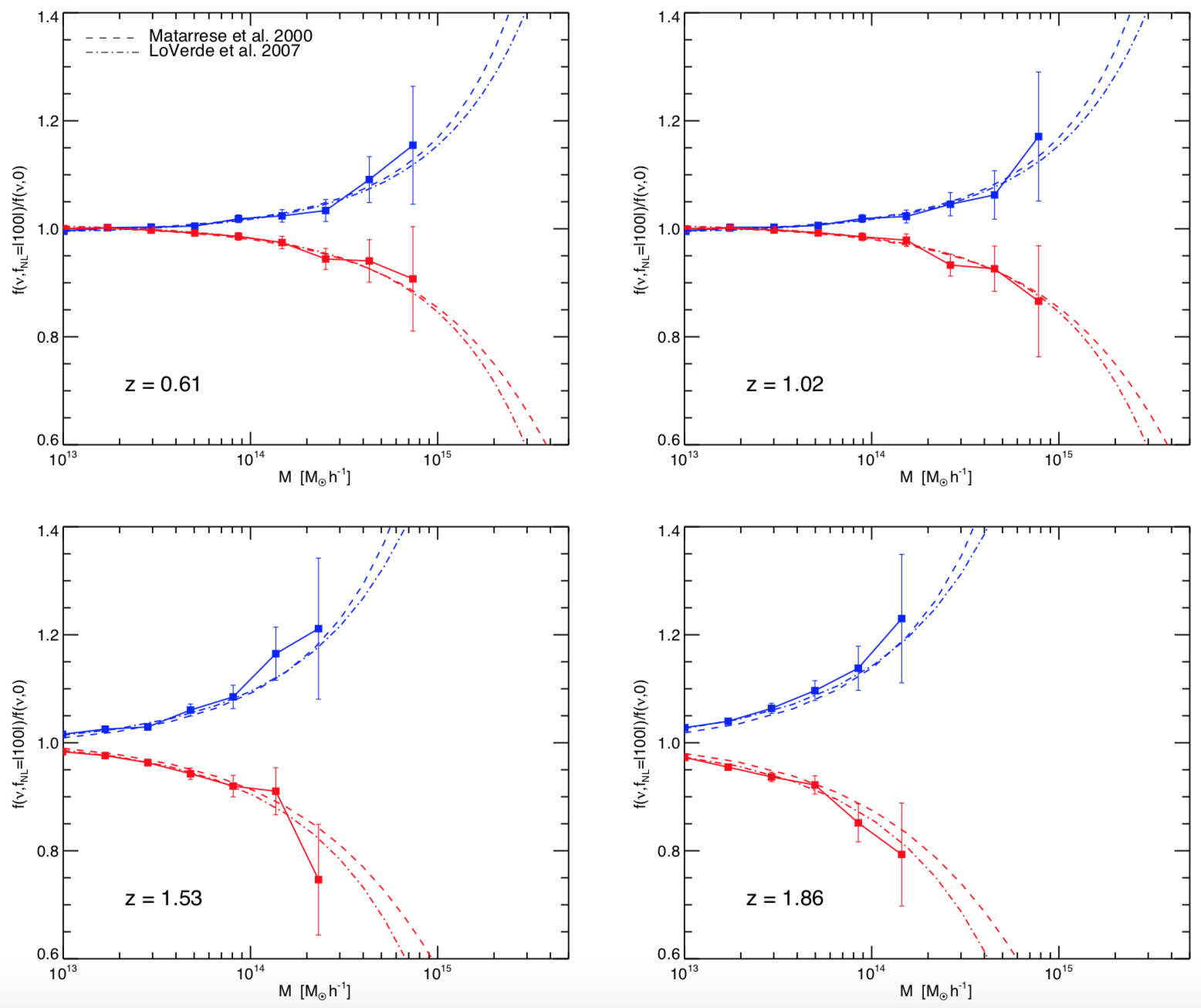}
    \caption{Ratio of the non-Gaussian ($f_{\rm NL} = \pm 100$) to Gaussian mass function for different redshift snapshots: top left $z = 0.61$; top right $z = 1.02$; bottom left $z = 1.53$; bottom right $z = 1.86$. The dashed line is the mass function of \cite{Matarrese:2000iz} and the dot-dashed lines are that of \cite{LoVerde:2007ri}, both including the $q$-correction. From \cite{Grossi:2009an}.}
\label{GrossiSimulationNG2}
\end{figure}

In \cite{Verde:2000vr, Matarrese:2000iz} it was shown that clusters at high redshift ($z>1$) can probe NG down to $f_{\rm NL} \sim 10^2$, see also \cite{LoVerde:2007ri} for an alternative approach. 
Actually, many methods have been developed for the  determination of mass function, such as
\begin{itemize}
\item the stochastic approach (first-crossing of a diffusive barrier)  \cite{Maggiore:2009rv, Maggiore:2009rw, Maggiore:2009rx, Maggiore:2009hp}, 
\item the ellipsoidal collapse method \cite{Lam:2008rk, Lam:2009nb}, 
\item  a combination of  saddle-point and diffusive barrier  \cite{DAmico:2010ywu}, 
\item the Log-Edgeworth expantion \cite{LoVerde:2011iz}, 
\item  the excursion sets studied with correlated steps  \cite{Paranjape:2011wa, Paranjape:2011ak}.
\end{itemize}
Remarkably,  excellent agreement of analytical formulae with N-body simulations (e.g. see Figure \ref{GrossiSimulationNG} for DM halos in NG simulations and  Figure \ref{GrossiSimulationNG2} for the ration of the non-Gaussian $f_{\rm NL}$ to Gaussian mass function) have been found in \cite{Pillepich:2008ka, Grossi:2009an, Desjacques:2008vf}, and in many other papers afterwards e.g. \cite{Sefusatti:2011gt, Lazanu:2015bqo, Lazanu:2015rta}.

Moreover, halo (galaxy) clustering and halo (galaxy) higher-order correlation functions represent further and more powerful implementations of this general idea. 
\\
Actually, the halo mass function (\`a-la-Press-Schechter) can be a useful tool to probe PNG as it essentially depends exponentially on the PNG parameters \cite{Matarrese:2000iz}, by modulating the critical overdensity for collapse. Its calculation can be done along the lines of the original Press-Schechter approach, using a steepest-descent approximation to deal with (small) PNG. 
However, several effects have to be carefully considered such as  non-Markovianity, already there in the Gaussian case, but unavoidable in NG case, or the details of the non-spherical collapse.
While analytical treatments are welcome, the validation with N-body simulations is  crucial.
Still we need to better understand the  connection between analytical and numerical quantities and real observables, and to what level is this affecting NG  measurements. 
 \begin{itemize}
 \item  Should we necessarily go on with (extended) Press-Schechter-like approaches? 
 \item Are alternative approaches viable such as Smoluchowski equation for the non-Poissonian random process (or the earliest attempt proposed in \cite{1978ApJ...223L..59S})?  
\end{itemize} 
In conclusion,  rare events (e.g. high-z and  massive clusters) offer interesting and promising opportunities for the detection of PNG, as both the mass-function of massive haloes and the number-counts of massive haloes are affected. 

\subsection{Halo bias in NG models}

As it is well known, halos (galaxies) do not trace the underlying (dark) matter distribution. For this reason, following the original proposal \cite{Kaiser:1984sw}, we introduce the ``bias''  parameters or Eulerian bias for galaxy clusters and later for galaxies  (for a review see \cite{Desjacques:2016bnm})
\begin{equation}
\delta_{halo}(x)=b_1\delta_{matter}(x)+b_2\delta_{matter}^2(x)+\dots
\end{equation}
that allow to parametrize our ignorance about the way in which dark matter halos cluster in space with respect to the underlying dark matter.
\\
Note that a complete set of the local bias terms (representing all possible local gravitational observables along the fluid trajectory) was presented in \cite{Assassi:2014fva, Senatore:2014eva, Mirbabayi:2014zca}.
For the most general expansion up to second order in the Eulerian framework with Gaussian initial condition see also \cite{1993MNRAS.262.1065C, Fry:1992vr, Fry:1996fg, Catelan:1997qw, Catelan:2000vn, McDonald:2009dh, Elia:2010en, Chan:2012jj, Baldauf:2012hs}.
The various bias parameters can be generally regarded either as purely phenomenological ones (i.e. to be fitted to observations) or predicted by a theory (e.g. Press-Schecter together with Lagrangian perturbation theory).
\\
Specifically, considering $\delta_{halo}(x)=b\ \delta_{matter}(x)$, it is possible to show that the halo  bias is sensitive to PNG through a scale-dependent correction term (in Fourier space), see e.g.  \cite{Dalal:2007cu, Slosar:2008hx, Matarrese:2008nc, Afshordi:2008ru, Verde:2009hy, Schmidt:2010gw, Desjacques:2010nn, Desjacques:2011mq, Schmidt:2012ys}. In particular, we have 
\begin{equation}
\frac{\delta b(k)}{b}\sim \frac{2 f_{\rm NL}\delta_c}{k^2}\, .
\end{equation}
This opens interesting prospects for 
constraining or measuring NG in LSS but 
demands for an accurate evaluation of the
effects of (general) NG on halo biasing.

The idea is to start from the results obtained in the 80’s in \cite{1986ApJ...310...19G, Matarrese:1986et} giving the general expression for the peak 2-point function as a function of N-point connected correlation functions of the background linear (i.e. Lagrangian) mass-density field 
\begin{equation}
\xi_{h,M}(|\mathbf{x}_1-\mathbf{x}_2|)=-1+\exp\left\{
\sum_{N=2}^\infty \sum_{n=1}^{N-1}
\frac{\nu ^N \sigma_R^{-N}}{j!(N-1)!}
\xi^{(N)}\left[\underset{\text{$j$ times}}{\mathbf{x}_1, \dots,  \mathbf{x}_1}, \underset{\text{$(N-j)$ times}}{\mathbf{x}_2, \dots ,  \mathbf{x}_2} \right]
\right\}
\end{equation}
which requires many techniques such as the path-integral, the cluster expansion, the multinomial theorem and asymptotic expansion. The analysis of NG models was motivated in \cite{1986Natur.323..132V} on bulk flows.

In \cite{Matarrese:2008nc} this relation was applied  to the case of NG of the gravitational potential, obtaining the power-spectrum of dark matter halos modeled as high ``peaks'' (up-crossing regions) of height $v=\delta_c/\sigma_R$ of the underlying mass density field (Kaiser's model). Here $\delta_c(z)$ is the critical overdensity for collapse (at redshift $z$) and $\sigma_R$ is the root mean square  mass fluctuation on scale $R$ ($M \sim R^3$).

The motion of peaks (going from Lagrangian to Eulerian space), which  implies \cite{Catelan:1997qw}
\begin{equation}
1+\delta_h(\mathbf{x}_{Eulerian})=(1+\delta_h(\mathbf{x}_{Lagrangian}))(1+\delta_R(\mathbf{x}_{Eulerian}))
\end{equation}
    and (to linear order) $b=1+b_L$ \cite{Mo:1995cs}, allows to derive the scale-dependent halo bias in the presence of NG initial conditions. Corrections may arise from second-order bias and GR terms.

\begin{figure}
\centering
\includegraphics[width=0.7\textwidth]{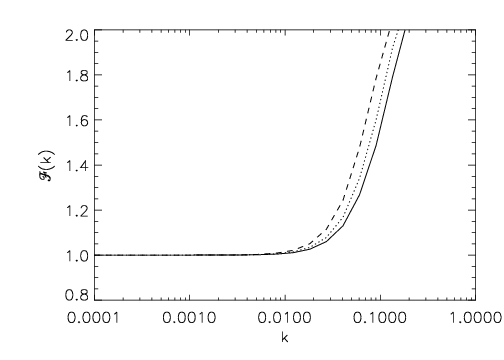}
\caption{The function $\mathcal{F}_R(k)$ for three different masses: $1\times 10^{14} M_{\odot}$ (solid), $2\times 10^{14} M_{\odot}$ (dotted), $1\times 10^{14} M_{\odot}$ (dashed). From \cite{Matarrese:2008nc}.}
\label{Fk_M_lin}
\end{figure}

Alternative approaches (e.g. based on 1-loop calculations) have been developed in \cite{Taruya:2007xy, Taruya:2008pg,  Matsubara:2008wx, Jeong:2009vd}. Improvements in the fit with N-body simulations by assuming dependence on gravitational potential have been carried out in \cite{Giannantonio:2009ak}, while for the extension to bispectrum see \cite{Baldauf:2010vn}.  Finally, the inclusion of $g_{\rm NL}$ and $f_{\rm NL}$ in analysis of QSO clustering was performed in \cite{Leistedt:2014zqa}.
\\
Note that  the extension to general (scale and configuration dependent) NG is
straightforward \cite{Matarrese:2008nc}.
Actually, we can write,  in full generality,   the $f$ bispectrum as $B_f(k_1,k_2,k_3)$. Then,  the relative 
NG correction to the halo bias is 
\begin{equation}
\begin{aligned}
\frac{\Delta b_h}{b_h}=&\frac{\Delta_c(z)}{D(z)}\frac{1}{8\pi^2\sigma_R^2} \int dk_1 \, k_1^2\mathcal{M}_R(k_1) \times
\\
&\int_{-1}^1d\mu \frac{\mathcal{M}_R(\sqrt{\alpha})}{\mathcal{M}_R(k)} \frac{B_{\phi}(k_1, \sqrt{\alpha}, k)}{P_{\phi}(k)}
\end{aligned}
\end{equation}
where $\alpha=k_1^2+k_2^2+2k_1k\mu$,  $P_\phi$ is the power-spectrum of a Gaussian gravitational potential, while $\mathcal{M}_R$ is the factor connecting the smoothed linear overdensity with the primordial potential by means of the factor
\begin{equation}
\mathcal{M}_R(k)=\frac{2}{3}\frac{T(k)k^2}{H_0^2\Omega_{m,0}}W_R(k)
\end{equation}
where $T(k)$ is the transfer function and $W_R(k)$ is the window function defining the radius $R$ of a 
proto-halo of mass $M(R)$.
It also applies to non-local (e.g. ``equilateral'') PNG (corresponding to DBI or  ghost 
inflation) and universal PNG term (see also \cite{Schmidt:2010gw, Scoccimarro:2011pz, Schmidt:2012ky, Assassi:2015fma}).
The halo bias in NG models has been calculated in \cite{Matarrese:2008nc}, the result is
\begin{equation}
b_h^{f_{\rm NL}}=1+\frac{\Delta_c(z)}{\sigma_R^2D^2(z)}\left[
1+2f_{\rm NL}\frac{\Delta_c(z)}{D(z)}\frac{\mathcal{F}_R(z)}{\mathcal{M}_R(z)}
\right]
\end{equation}
where the form factor is given by
\begin{equation}
\mathcal{F}_R(k)=\frac{1}{8\pi^2\sigma_R^2}\int dk_1k_1^2\mathcal{M}_{R}(k_1) P_\phi(k_1)\int_{-1}^1 d\mu \mathcal{M}_R(\sqrt{\alpha}) \left[
\frac{P_\phi(\sqrt{\alpha)}}{P_\phi(k)}+2
\right]
\end{equation}
and plotted, for three different masses,  in Figure \ref{Fk_M_lin}.

\subsection{PNG with LSS: the galaxy bispectrum}

The bispectrum of galaxies can be used  to forecast the constraining power of LSS surveys on measuring the amplitude of PNG, see e.g. \cite{Scoccimarro:2003wn, Sefusatti:2007ih}. 
In this section, we derive, following \cite{Tellarini:2016sgp}, the  galaxy bispectrum.
\\
The starting point is the relation between the linearly evolving density field $\delta_{lin}$ and the primordial gravitational potential 
\begin{equation}
\delta_{lin}(\mathbf{k},z)=\alpha(k,z)\Phi_{in}(\mathbf{k})
\end{equation}
with $\alpha(k,z)$  defined as
\begin{equation}
\alpha(k,z)\equiv \frac{2k^2c^2T(k)D(z)}{3\Omega_mH_0^2}
\end{equation}
where $T(k)$ is the transfer function. 
Since the linearly evolving density includes non-Gaussian terms in presence of PNG, we can  define the Gaussian part as
\begin{equation}
\delta_G(\mathbf{k}, z)=\alpha(k,z)\varphi_G(\mathbf{k}) \, .
\end{equation}
With this definition, using second-order (Eulerian) perturbation theory, we get \cite{Bernardeau:2001qr}
\begin{equation}\label{delta2}
\begin{aligned}
\delta^{(2)}(\mathbf{k}, z)&=\int \frac{d\mathbf{k}_1}{(2\pi)^3}\int \frac{d\mathbf{k}_2}{(2\pi)^3}
\delta^D(\mathbf{k}-\mathbf{k}_1-\mathbf{k}_2)\times
\\
&\times\left[
\mathcal{F}_2(\mathbf{k}_1, \mathbf{k}_2)+ f_{\rm NL}\frac{\alpha(k)}{\alpha(k_1)\alpha(k_2)}
\right]\delta_G(\mathbf{k}_1, z) \delta_G(\mathbf{k}_2, z) 
\end{aligned}
\end{equation}
where the 2-nd order gravity-kernel  $\mathcal{F}_2$ is given by 
\begin{equation}
\mathcal{F}_2(\mathbf{k}_1, \mathbf{k}_2)=\frac{5}{7}+\frac{1}{2}\frac{\mathbf{k}_1\cdot \mathbf{k}_2}{k_1k_2}\left(
\frac{k_1}{k_2}+\frac{k_2}{k_1}
\right)+\frac{2}{7}\frac{(\mathbf{k}_1\cdot \mathbf{k}_2)^2}{k_1^2k_2^2} \, .
\end{equation}
The quantity $\delta(x)$ is expressed in the Eulerian frame, with the initial spatial coordinate $\mathbf{q}$ in the Lagrangian frame being related to the evolved Eulerian coordiate $\mathbf{x}$ through the formula
\begin{equation}
\mathbf{x}(\mathbf{q}, \tau)=\mathbf{q} + \mathbf{\Psi} (\mathbf{q}, \tau)
\end{equation} 
where $\mathbf{\Psi} $ is the displacement field. 
Using this relation, we can rewrite the second-order solution appearing in (\ref{delta2}) as \cite{peebles1980large, Bouchet:1992uh}
\begin{equation}
\delta^{(2)}(\mathbf{x}, \tau)=\frac{17}{21}(\delta_{lin}(\mathbf{x}, z))^2+\frac{2}{7}s^2(\mathbf{x}, z)
-
\mathbf{\Psi}(\mathbf{x}, z)\cdot \nabla\delta(\mathbf{x}, z)
\end{equation} 
where $s^2=s_{ij}s^{ij}$ and $s_{ij}$  is the trace-free tidal tensor, defined as
\begin{equation}
s_{ij}\equiv \left(
\nabla_i\nabla_j - \frac{1}{3}\delta_{ij}^K\nabla^2) \nabla^{-2}\delta
\right)
\end{equation}
and $\delta_{ij}^K$ is the Kronecker delta. In the following, we will omit the redshift dependence in the density and velocity fields.

 We can introduce a long-short splitting of the gravitational potential and DM density field, such that, for local NG one  easily finds (for local NG)
 \begin{equation}
 \delta_{lin,\ell}(\mathbf{k})=\delta_{G,\ell}+f_{\rm NL}\alpha(\varphi_{G,\ell}^2-\left<\varphi_{G,\ell}^2\right>)
 \end{equation}
 with
 \begin{equation}
 \varphi_G(\mathbf{q})= \varphi_{G,\ell}(\mathbf{q})+ \varphi_{G,s}(\mathbf{q}) \, .
 \end{equation}
 In Lagrangian space one can then introduce the expansion 
 \begin{equation}
 \begin{aligned}
 \delta_g^L(\mathbf{q})&=\frac{n_g(\mathbf{q})-\left<n_g\right>}{\left<n_h\right>}
 \\
 &=
 b_{10}^L\delta_{lin}+b_{01}^L\varphi_G+b_{20}^L(\delta_{lin})^2+b_{11}^L\delta_{lin}\varphi_G+b_{02}^L\varphi_G^2+\dots
 \end{aligned}
 \end{equation}
 where  the $5$ $b_{ij}$ represent our (generally unknown) bias parameters.
 \\
The final Eulerian position of the galaxy can be obtained by using the conservation 
law \cite{Catelan:1997qw}
\begin{equation}
1+\delta_g^E(\mathbf{x},z)=[1+\delta(\mathbf{x},z)][1+\delta_g^L(\mathbf{q},z)]
\end{equation}
to find
\begin{equation}
\delta_g^E(\mathbf{k})=\delta_{10}^E\delta+b_{01}^E\varphi_G+b_{20}^E\delta\ast \delta
+b_{11}^E\delta\ast \varphi_G+b_{02}^E\varphi_G\ast\varphi_G
-\frac{2}{7}b_{10}^Ls^2-b_{01}n^2
\end{equation}
where $s^2$ and $n^2$ are suitable expansion terms and the Eulerian bias parameters read:
\begin{equation}
\begin{aligned}
b_{10}^E&=1+b_{10}^L\,,
\\
b_{01}^E&=b_{01}^L\,,
\\
b_{20}^E&=\frac{8}{21}b_{10}^L+b_{20}^L\,,
\\
b_{11}^E&=b_{01}^L+b_{11}^L\,,
\\
b_{20}^E&=b_{02}^L \, .
\end{aligned}
\end{equation}
Using the  standard definitions for the galaxy power spectrum $P_{gg}$ and bispectrum $B_{ggg}$
\begin{equation}
\begin{aligned}
\left< 
\delta_g^E(\mathbf{k}_1)\delta_g^E(\mathbf{k}_2)
\right>
&=
(2\pi)^3\delta^D(\mathbf{k}_1+\mathbf{k}_2) P_{gg}(\mathbf{k}_1) \, ,
\\
\left<
\delta_g^E(\mathbf{k}_1)\delta_g^E(\mathbf{k}_2)\delta_g^E(\mathbf{k}_3)
\right>
&=
(2\pi)^3\delta^D(\mathbf{k}_1+\mathbf{k}_2+\mathbf{k}_3)B_{ggg}(\mathbf{k}_1,\mathbf{k}_2,\mathbf{k}_3) \, .
\end{aligned}
\end{equation}
we can simply write at tree-level
\begin{equation}
\begin{aligned}
P_{gg}(\mathbf{k}_1)&=E_1^2(\mathbf{k}_1)P(k_1)
\\
B_{ggg}(\mathbf{k}_1,\mathbf{k}_2,\mathbf{k}_3)&=
2E_1(\mathbf{k}_1)E_1(\mathbf{k}_2)E_2(\mathbf{k}_1, \mathbf{k}_2) P(k_1)P(k_2)+2\  \text{cyc.}
\end{aligned}
\end{equation}
where $P(k)$ is the matter power spectrum for the Gaussian source field $\varphi_G$, while the kernels $E_i$ are defined as
\begin{equation}
E_1(\mathbf{k}_1)=b_{10}+\frac{b_{01}}{\alpha(k_1)}
\end{equation}
with the scale dependent bias term $b_{01}/\alpha(k_1) \propto f_{\rm NL}/k_1^2 $, and
\begin{equation}
\begin{aligned}
E_2(\mathbf{k}_1,\mathbf{k}_2)=&
\ b_{10}\left[
F_2(\mathbf{k}_1,\mathbf{k}_2)+f_{\rm NL}\frac{\alpha(|\mathbf{k}_1+\mathbf{k}_2|)}{\alpha(k_1)\alpha(k_2)}
\right]
\\
&+
\left[
b_{20}-\frac{2}{7}b_{10}^LS_2(\mathbf{k}_1,\mathbf{k}_2)
\right]
+\frac{b_{11}}{2}\left[
\frac{1}{\alpha(k_1)}+\frac{1}{\alpha(k_2)}
\right]
\\
&+\frac{b_{02}}{\alpha(k_1)\alpha(k_2)}
-b_{01}\left[
\frac{N_2(\mathbf{k}_1, \mathbf{k}_2)}{\alpha(k_2)}+\frac{N_2(\mathbf{k}_2, \mathbf{k}_a)}{\alpha(k_1)}
\right]\, .
\end{aligned}
\end{equation}
Note that, general relativistic effects (including also redshift-space distortions, lensing, etc.) have to be taken into account both in the galaxy power-spectrum and bispectrum, as well as in the dark matter evolution.
\\ 
The complete expression (very involved) for the galaxy bispectrum was written  down for the first time recently, and can be found in \cite{Bertacca:2017dzm} to be soon compared with observations, see also \cite{DiDio:2015bua, Raccanelli:2015vla, Umeh:2016nuh}.

We conclude this section  describing, following  \cite{Tellarini:2016sgp},  the Fisher matrix forecasts   on $\sigma_{f_{\rm NL}}$ (the accuracy of the determination of local non-linear parameter $f_{\rm NL}$) from measurements of the galaxy bispectrum, as well as the constraints on PNG  from the galaxy power spectrum and bispectrum  in future radio continuum and optical surveys  \cite{Karagiannis:2018jdt}.  See also \cite{Giannantonio:2011ya} for the constraining power on primordial non-Gaussianity of  the Dark Energy Survey  (DES)  as well as  Euclid and WFIRST.
\\
Particularly, the tree-level bispectra with local non-Gaussian initial conditions are shown in Table \ref{tab:others_fnl}  in redshift space, where the covariance between different triangles has been neglected.
\\
While many  issues are still present (such as the need for full covariance, a better understanding of accurate bias model, the inclusion of general relativistic effects, or the proper implementation of the estimators), still moving from the 2D {\it Planck} data to the 3D maps of the forthcoming surveys represent a  great potential as
\begin{itemize}
\item  the bispectrum could do better than the power-spectrum,
\item we might  increase the accuracy to achieve $f_{\rm NL} \sim 1$.
\end{itemize}
In particular, the LSS bispectrum allows in principle tight constraints also on non-local   shapes (e.g. equilateral).
However, even if naive mode counting suggest that $\sigma_{f_{\rm NL}} \sim 1$ for the equilateral shape might be achievable by     pushing $k_{max}$ high enough, modeling  the gravitational bispectrum  in the non-linear regime  with high accuracy is very challenging, as the equilateral shape is more correlated than local to the non-linear gravitational bispectrum.

   \begin{table}[ht]
\centering
\begin{tabular}{cccccc}
\hline
		      &  \multicolumn{2}{c}{Power Spectrum}	&   \multicolumn{2}{c}{Bispectrum} \\
Sample		      & $\sigma_{f_{\rm NL}}$ & $\sigma_{f_{\rm NL}}$ 	& $\sigma_{f_{\rm NL}}$ 	  		  & $\sigma_{f_{\rm NL}}$  			\\
		      & bias float	    & bias fixed	& bias float		  		  & bias fixed 				\\  
\hline
BOSS 	  	      &  $21.30$	    & $13.28$	        & $ 1.04^{(0.65)}_{(2.47)} $		  &  $ 0.57^{(0.35)}_{(1.48)} $        \\  
eBOSS          &  $14.21$        & $11.12$          & $ 1.18^{(0.82)}_{(2.02)} $          &  $ 0.70^{(0.48)}_{(1.29)} $        
\\
Euclid 	  	  &  $6.00$	    &  $4.71$		& $ 0.45^{(0.18)}_{(0.71)} $		  &  $ 0.32^{(0.12)}_{(0.35)} $        \\   
DESI  		  &  $5.43$	    &  $4.37$		& $ 0.31^{(0.17)}_{(0.48)} $		  &  $ 0.21^{(0.12)}_{(0.37)} $        \\   
\hline
BOSS + Euclid     &  $5.64$	    &  $4.44$		&$ 0.39^{(0.17)}_{(0.59)} $	   	  &  $ 0.28^{(0.11)}_{(0.34)} $        \\  
\hline\hline  
\end{tabular}
\caption{Forecasts for $\sigma_{f_{\rm NL}}$, the accuracy of the determination of local $f_{\rm NL}$, from the bispectrum of BOSS, eBOSS, DESI and Euclid. From  \cite{Tellarini:2016sgp}.}    
\label{tab:others_fnl}
\end{table} 

Finally,  possible future constraints on the amplitude $f_{\rm NL}$ for the local, equilateral and orthogonal  shapes, through galaxy power spectrum and bispectrum measurements on large scales based on radio continuum (with 10 $\mu$Jy and  1 $\mu$Jy flux limits) and optical (spectroscopic and photometric) surveys are presented in Table \ref{table:concl}, see \cite{Karagiannis:2018jdt} for details.
Remarkably, for the local shape, LSS measurements can provide significant improvements over current {\it Planck} constraints on PNG.  

   \begin{table}[ht]
\centering
\begin{tabular}{cccccc}
\hline
                                 & \Planck  &  1 $\mu$Jy     &  10 $\mu$Jy  &  Spectroscopic  & Photometric       \\ \hline

\multicolumn{1}{c|}{Local}           &  5.0  & 0.2 & 0.6  & 1.3       & 0.3             \\ 
\multicolumn{1}{c|}{Equilateral}   & 43    & 244  & 274   & 57   & 184 \\ 
\multicolumn{1}{c|}{Orthogonal}  & 21    & 18    & 29  & 18     & 38      \\ \hline
\end{tabular}
\caption{Summary of $1\sigma$ limits for the three PNG types considered, from radio continuum and optical surveys  derived from combining the power spectrum and bispectrum and accounting for RSD, the trispectrum term and theoretical errors. From \cite{Karagiannis:2018jdt}.}
\label{table:concl}
\end{table}

\section{Controversial issues on non-Gaussianity}
In this section, we present some  aspects regarding  non-Gaussianity which we think are still controversial or require further clarifications. 
\subsection{Single-field consistency relation}
The first issue concerns the single-field consistency relation. Actually, as mentioned before, the common lore is that the  ``consistency relation'', implying $f_{\rm NL} = - 5\, (n_s-1)/12$,  can be gauged away by a non-linear rescaling of coordinates, up to sub-leading terms. 
As a consequence,  the only residual term is proportional to $\epsilon$ thus of the same order of the amplitude of tensor modes. 
\\
More precisely, the bispectrum for single-field inflation  can be represented as \cite{Gangui:1993tt, Acquaviva:2002ud, Maldacena:2002vr}:
\begin{equation}
\begin{aligned}
B_\zeta(k_1, k_2, k_3) &\propto \frac{(\Delta_\zeta^2)^2}{(k_1k_2k_3)^2}\left[ 
(1-n_s)\mathcal{S}_{loc.}(k_1, k_2, k_3)+\frac{5}{3}\epsilon \mathcal{S}_{equil.}(k_1, k_2, k_3)
\right]
\\
n_s&=1-\eta-2\epsilon \,, \quad \text{with} \quad \epsilon\equiv \frac{\dot{H}}{H^2} \, , \ \eta\equiv \frac{\dot{\epsilon} }{H\epsilon}\, .
\end{aligned}
\end{equation}
The observability of the so-called ``Maldacena consistency relation'', related to the above bispectrum for single field inflation, in CMB and LSS data, has led to a long-standing controversy. 
\\
Recently, various groups have argued that the $(1-n_s)$ term is totally unobservable (for single-clock inflation), as, in the strictly squeezed limit (one of the wave-numbers, say $k_i$, going to 0, $k_i\rightarrow 0$), this term can be gauged away by a suitable coordinate tranformation. 
\\
However, in \cite{Cabass:2016cgp} it has been  argued that the term survives up to a ``renormalization'' which further reduces it by a factor of $\sim  0.1$ if one applies Conformal Fermi Coordinates (CFC) to get rid of such a ``gauge mode''.
\begin{itemize}
\item Is this (CFC approach) the only way to deal with this term? 
\item Can we aim at an exact description, which is not affected by ``spurious PNG''?
\end{itemize}

\subsection{Non-Gaussian $f_{\rm NL}$-like terms generated by non-linear general relativistic evolution}
An other important issue regards the role of the non-linear  evolution of the matter perturbations in general relativity.
Actually,  second order DM dynamics in GR leads to (post-Newtonian) $\delta\zeta$-like terms which mimic local primordial non-Gaussianity \cite{Bartolo:2005xa}.
For instance, these terms have been  included in the  halo bias in \cite{Verde:2009hy}. 
\\
 For a recent estimate of the effective non-Gaussianity due to general-relativistic lightcone effects mimicking a PNG signal, see \cite{Koyama:2018ttg} \footnote{Note that also dark energy could  in principle introduce  degeneracies with PNG, see e.g. \cite{Sefusatti:2006eu, Hashim:2018dek}.}.

Remarkably, these   GR terms can be  recovered by a short-long mode splitting ($\lambda_S$ and $\lambda_L$, respectively) leading to a resummed non-linear contribution $\delta e^{-2\zeta}$ \cite{Bruni:2014xma}. 
This comes from the modulation of sub-horizon scales due to modes entering the horizon at any given time.
In the comoving gauge (suitable for calculation of halo bias) this would correspond to an $f_{\rm NL} = -5/3$ in the pure squeezed limit. 
\\
Then, we may ask the following  question
\begin{itemize}
\item is such  relativistic NG signature detectable via some cosmological observables?
\end{itemize}  

\begin{figure}
  \centering
    \includegraphics[width=0.5\textwidth]{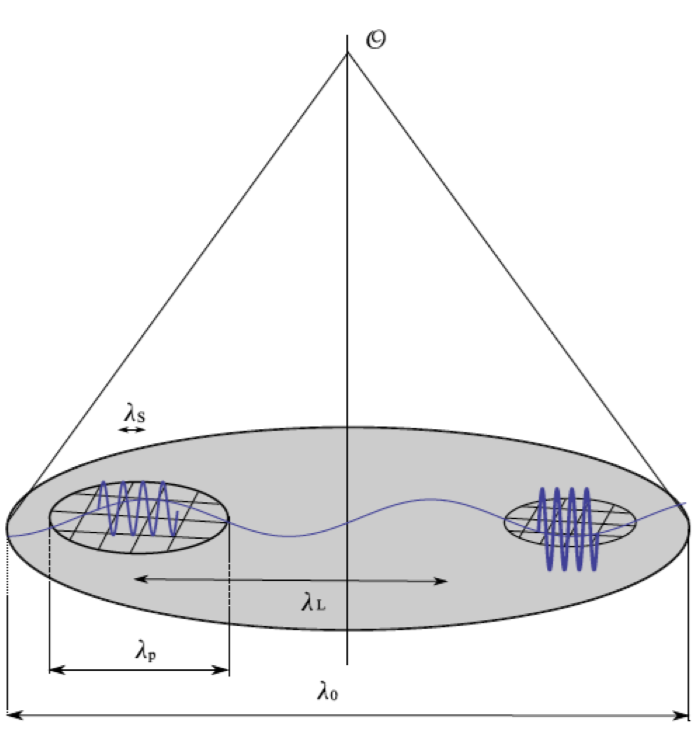}
    \caption{Schematic of the various scales in Eq. (\ref{lambdascales}). From \cite{Bartolo:2015qva}. }
\label{picture}
\end{figure}

Consider a patch of the Universe, where the comoving spatial element is given by
\begin{equation}
ds^2_{(3)} = e^{2\zeta} \delta_{ij} dx^i dx^j  \, .
\end{equation}
There is a global background which must be defined with respect to some scale $\lambda_0$, at least as large as all the other scales of interest, i.e., at least as large as our presently observable Universe. 
\\
Then, there is an other important scale, the separate Universe patches, $\lambda_P$, distinguished from $\lambda_0$ and we assume $\lambda_0\gg \lambda_P\gg \lambda_S $. This is large enough for each patch to be treated as locally homogeneous and isotropic, but  patches must be stitched together to describe the long-wavelength perturbations on a scale $\lambda_L \gg \lambda_P $, see Figure \ref{picture}. Thus, 
\begin{equation}
\label{lambdascales}
\lambda_0> \lambda_L \gg \lambda_P \gg \lambda_S  \, .
\end{equation}

The local observer in a separate Universe patch cannot observe the effect of $\zeta_L$, which is locally homogeneous on the patch scale $\lambda_P$. However, local coordinates can be defined only locally and the long mode curvature perturbation is observable through a mapping from local to global coordinates.

In the halo bias case the effect is unobservable. Indeed, as pointed out in  \cite{Dai:2015rda, Dai:2015jaa, dePutter:2015vga}, a local physical redefinition of the mass gauges away such a NG effect (in the pure squeezed limit), similarly to Maldacena's  single-field NG contribution.
This is true provided the halo bias definition is strictly local. 
We may ask the following  questions:
\begin{itemize}
\item are there significant exceptions? 
\item are all non-linear GR effects fully accounted for by ``projection effects''?
\end{itemize}
In general, this dynamically generated GR non-linearity is physical and cannot be gauged away by any local mass-rescaling, provided it involves scales larger than the patch required to define halo bias, but smaller than the separation between halos (and the distance of the halo to the observer). 
\\
Hence one would expect it to be in principle detectable in the matter bispectrum. Similarly, the observed galaxy bispectrum obtained via a full GR calculation must include all second-order GR non-linearities on such scales (only as projection effects?).

In conclusion, the separate Universe approach is very useful for many applications, but the effect of the external world cannot be always described by linear theory, thus the usual identification of large scales with the linear theory is only qualitative and can become misleading in some cases.
\\
 For example, on one hand, perturbations of order $N \gg 1$ give the leading contribution to $N$-th order moments, such as $\left<\delta^N\right>_c$. 
On the other hand, we know from non-linear Newtonian dynamics  that  $\left<\delta^N\right>_c \sim  \left<\delta^2\right>^{N-1}$ on all scales (for scale-free spectra).
Well inside a given separate Universe, the  assumption  that the only non-linearity is described by Newtonian physics might be too restrictive, as the relevance of non-linear GR effects in sub-patch dynamics depends upon the specific problem.

It would be interesting to see the effects of using, for example,   the silent Universe description (\cite{Bruni:1994nf}) to account for deviations of the patch from purely spherical behavior.
Recall   that over-dense patches evolve towards oblate ellipsoids and even under-dense ones can collapse to oblate ellipsoids, owing to tidal effects of surrounding matter. 
Recent approaches using the local tide approximation \cite{Ip:2016jji} go in this direction.

\section{Concluding remarks}
The concordance  $\Lambda$CDM cosmological model  describes the evolution of the Universe from 380,000  years after the Big Bang to the present time with astonishingly accuracy.
However, the mechanism that generated the primordial fluctuations  representing the seeds of the structures we observe in the CMB and in the LSS is unknown.
Remarkably, inflation provides a causal mechanism for the generation of cosmological perturbations, whose detailed predictions are fully supported by CMB and LSS data.
\\
Clearly, the direct detection of  PGW  and PNG   with the specific features predicted by inflation would provide strong independent support to this framework.

In the previous sections, we have summarized the theoretical motivations for PNG and the present observational status.
As stated before, {\it Planck} has provided stringent limits on $f_{\rm NL}$ which will be improved with polarization maps and using the full data in the upcoming  ``{\it Planck} legacy'' paper.
\\
In conclusion, among the short term goals, we need to look for more non-Gaussian shapes, such as scale-dependent $f_{\rm NL}$, make use of the bispectrum in 3D data and improve the constraints on $g_{\rm NL}$.

The final goal is to reconstruct the inflationary action by improving the sensitivity on NG parameters,  searching for $f_{\rm NL} \sim 1$ for all shapes, taking into account non-linear general relativistic effects and second-order radiation transfer function contributions. 

\acknowledgments

SM would like to thank the directors of the school, Eugenio Coccia, Joe Silk and Nicola Vittorio, and the Italian Physical Society (SIF) for having invited him to lecture at the International School of Physics Enrico Fermi (from 26th June to 19th July 2017) in the beautiful Villa Monastero, Varenna, Lake Como (Italy). SM acknowledges partial financial support by ASI Grant No. 2016-24-H.0.

\bibliographystyle{varenna}
\bibliography{references_NG}

\end{document}